\documentclass[
 aps,
 prr,
 twocolumn,
 floats,
 epsf,
 amsmath,amssymb,
]{revtex4-2}

\usepackage{graphicx}
\usepackage{dcolumn}
\usepackage{bm}
\usepackage{hyperref}
\usepackage[capitalise]{cleveref}
\usepackage{bbold}
\usepackage{MnSymbol}
\usepackage{physics}
\usepackage{extarrows}
\usepackage[normalem]{ulem}
\usepackage{placeins}
\usepackage{booktabs}

\usepackage{colortbl}
\usepackage{xcolor}
\usepackage{microtype}
\usepackage{cleveref}
\crefrangeformat{equation}{Eqs.~(#3#1#4) to (#5#2#6)}
\crefformat{equation}{Eq.~(#2#1#3)}
\crefmultiformat{equation}{Eqs.~(#2#1#3)}
{,~(#2#1#3)}{,(#2#1#3)}{,~(#2#1#3)}
\crefformat{table}{Tab.~#2#1#3}
\crefmultiformat{table}{Tabs.~#2#1#3}
{,~#2#1#3}{,#2#1#3}{,~#2#1#3}
\crefrangeformat{table}{Tabs.~#3#1#4 to #5#2#6}
\crefformat{figure}{Fig.~#2#1#3}
\crefmultiformat{figure}{Figs.~#2#1#3}
{,~#2#1#3}{,#2#1#3}{~#2#1#3}
\crefrangeformat{figure}{Figs.~#3#1#4 to #5#2#6}

\newcommand{\TUVienna}
{\affiliation{$^a$Institute of Solid State Physics, TU Wien, 1040 Vienna, Austria}}
\newcommand{\UniWuerz}
{\affiliation{$^b$Institut f\"ur Theoretische Physik und Astrophysik and W\"urzburg-Dresden Cluster of Excellence ct.qmat, Universit\"at W\"urzburg, 97074 W\"urzburg, Germany}}

\begin{document}

\preprint{}

\title{General Shiba mapping for on-site four-point correlation functions
}

\author{Herbert E{\ss}l$\, ^{a}$}\email{herbert.essl@tuwien.ac.at}
\author{Matthias Reitner$\, ^a$}
\author{Giorgio Sangiovanni$\, ^b$}
\author{Alessandro Toschi$^a$}

 \TUVienna
 \UniWuerz

\date{\today}

\begin{abstract}
By applying the Shiba mapping on the two particle level, we derive the relation between the local four-point correlation functions of bipartite lattice models with on-site electronic repulsion and those of the corresponding models with attractive interaction in the most general setting. In particular, we extend the results of [Phys.~Rev.~B, {\bf 101}, 155148 (2020)], which were limited to the rather specific situation of the static limit in strictly particle-hole symmetric models, (i) by explicitly including {\sl both} magnetic field and different values of the chemical potentials, and (ii) by considering the full dependence of the generalized susceptibilities on the transfer (bosonic) Matsubara frequency. The derived formalism is then applied, as a relevant benchmark, to the Hubbard atom, by investigating the general properties of the divergences of its irreducible vertex functions as a function of chemical potential and 
applied magnetic field. The resulting phase diagrams provide an insightful compass for future studies of the breakdown of the self-consistent perturbation expansion beyond high-symmetric regimes.

\end{abstract}

\maketitle

\section{Introduction}
\label{sec:intro}

Due to the impressive algorithmic and computational advancements of the last decade, performing quantum-many body calculations of four point correlation functions has become achievable also in non-trivial, strongly correlated parameter regimes. Evidently, this progress is highly relevant for several reasons; among which we just recall here the precise calculations of vertex corrections of physical susceptibilities \cite{liu2012,hausoel2017,vilardi2018,watzenbock2020,watzenbock2022} and conductivities \cite{kauch2020,vranic2020,vucicevic2023}, the identification of the predominant scattering mechanisms underlying intriguing photoemission or self-energy features \cite{gunnarsson2015,wu2017,rohringer2020,schafer2021,wu2022,dong2022}, and the implementation of advanced diagrammatic expansions \cite{rohringer2018,krien2019,krien2020,delre2019,delre2021} for non-perturbative regimes.

On the theoretical side, this development has triggered a quest for improving our fundamental understanding of the many-electron properties on the two-particle level and their associated Feynman diagrammatic formalism.
In this context, a considerable effort  has been recently devoted to the analysis of generalized susceptibilities and two-particle vertex functions \cite{rohringer2012,hafermann2014,thunstrom2018,kugler2021,ge2024} beyond the standard textbook discussions, with a particular focus on their high-frequency asymptotics \cite{kunes2011,tagliavini2018,wentzell2020} and on their algorithmic treatments \cite{lee2021}.
Considering the specific case of the physical interpretation of the local (on-site) two-particle formalism, interesting information was obtained in Ref.~\cite{springer2020}. The Shiba transformation, which maps the quantities of a model with local electrostatic repulsion $U > 0$ to those of a corresponding Hamiltonian with on-site attraction $U < 0$, was applied to derive the explicit relations between the on-site generalized two-particle susceptibilities of the repulsive and their counterparts in the attractive Hubbard model. In fact, the analytical derivations presented in \cite{springer2020} unveiled intrinsic symmetry  properties of the four-point correlation functions (such as two-particle generalized susceptibilities and vertex functions) and provided pivotal information for the investigation of the breakdown of the self-consistent perturbation expansion in many-electron systems.

However, the analytical derivations of Ref.~\cite{springer2020} were limited to a very specific case: the Shiba transformation of the static (i.e., for zero frequency transfer, $\omega=0$) two-particle generalized susceptibilities in the particle-hole symmetric Hubbard model, i.e.~half filling $(\mu = U/2)$ {\sl and} $SU(2)$ symmetry $(h=0)$. 
In this paper, our goal is to address and overcome these notable restrictions, by providing  
the general expressions which systematically map all two-particle generalized on-site susceptibilities of a model with a purely local (Hubbard-like) interaction onto their counterparts in the corresponding model with sign-flipped interaction. Importantly, these mappings are established for arbitrary values of the chemical potential, applied magnetic field, and bosonic Matsubara frequency $\omega$.

Eventually, to demonstrate the correctness and usefulness of the derived expressions, we validate them through a hands-on application. Specifically, we exploit these expressions to analyze the divergences affecting the irreducible two-particle vertex functions of the Hubbard atom (i.e., of an isolated interacting site) across different scattering sectors, as a function of arbitrary chemical potential, magnetic field, temperature and interaction.
The obtained general phase diagrams will provide an intuitive and robust guidance to all future studies aiming at unveiling the formal \cite{schafer2013,kozik2015,gunnarsson2017,vucicevic2018} and the physical \cite{gunnarsson2016,reitner2020,chalupa2021,adler2024} aspects associated to the breakdown \cite{schafer2013,kozik2015,gunnarsson2017} of the self-consistent perturbation theory for many-electron systems, {\sl beyond} the (unrealistic) assumption of perfect particle-hole symmetry made so far in the largest part of the corresponding literature (with the notable exceptions of Refs.~\cite{gunnarsson2016,vucicevic2018, reitner2020}). \\

The structure of the paper is the following: In \cref{sec:formalism}, 
we concisely introduce the  one- and two-particle formalism necessary for our derivations, and we recall the general definition of the Shiba transformation. In \cref{sec:der_mapp}, we report the explicit derivation for the mapping of generalized on-site two-particle quantities under the action of the Shiba transformation in the most general case (i.e.~arbitrary filling, magnetic field, and finite transfer frequencies). Thereafter, in \cref{sec:application} we apply the derived relations to the problem of the divergences of the irreducible vertex functions occurring in the different scattering channels of the Hubbard atom with repulsive and attractive interaction, whose specific study was hitherto restricted to the case of  perfect particle-hole symmetry. Finally, in \cref{sec:conclusio} we draw our conclusions, outlining possible consequences of the results presented in our work.

\section{Formalism}
\label{sec:formalism}

\subsection{Generalized Susceptibilities}
\label{sec:gen_sus}

When examining the fermionic four-point correlation functions of systems that maintain time translation invariance, the associated generalized susceptibilities, which characterize the scattering events between two particles, depend only on three independent Matsubara frequencies (instead of four). This reduction is due to the conservation of energy~\cite{rohringer2012,rohringer2013}. Additionally, for spin conservation, only $6$ out of the $2^4$ spin combinations for these correlation functions remain independent~\cite{rohringer2012}.

Therefore, for our considerations all possible (local) generalized susceptibilities $\chi$ can be defined as follows~\cite{rohringer2018, rohringer2012}:

\begin{align}
\begin{split}
    \chi_{\sigma\sigma^\prime}(\nu_1,\nu_2,\nu_3)&=\int_0^\beta  d\tau_1 d\tau_2 d\tau_3 \, e^{-i\nu_1\tau_1}e^{i\nu_2\tau_2}e^{-i\nu_3\tau_3} \\
    &\bigl(\langle T_\tau c_\sigma^\dagger(\tau_1)c_\sigma(\tau_2)c_{\sigma^\prime}^\dagger(\tau_3)c_{\sigma^\prime}(0) \rangle \\
    -\langle T_\tau &c_\sigma^\dagger(\tau_1)c_\sigma(\tau_2) \rangle \langle T_\tau c_{\sigma^\prime}^\dagger(\tau_3)c_{\sigma^\prime}(0) \rangle\bigr),
\end{split} \label{eq:gensus}\\
\begin{split}
    \chi_{\overline{\sigma\sigma^\prime}}(\nu_1,\nu_2,\nu_3)&=\int_0^\beta  d\tau_1 d\tau_2 d\tau_3 \, e^{-i\nu_1\tau_1}e^{i\nu_2\tau_2}e^{-i\nu_3\tau_3} \\
    &\langle T_\tau c_\sigma^\dagger(\tau_1)c_{\sigma^\prime}(\tau_2)c_{\sigma^\prime}^\dagger(\tau_3)c_{\sigma}(0) \rangle,
\end{split} \label{eq:gensus_bar}
\end{align}
where $\sigma, \sigma'$ denote the spins, $\nu_i$ are fermionic Matsubara frequencies (where the fourth frequency is set by $\nu_4=\nu_1-\nu_2+\nu_3$), $c^{(\dagger)}_{\sigma}$ annihilate (create) an electron with spin $\sigma$, $T_\tau$ denotes the imaginary time ordering operator, and $\langle \dots \rangle = \frac{1}{Z}\Tr(e^{-\beta H} \dots)$ the thermal expectation value.

Note that for non-local $\chi$, $c^{(\dagger)}_{i,\sigma}$ acquire an additional lattice index $i$, and for space translation invariant systems, one of the four lattice indices can be set to zero.

By using two fermionic $\nu$, $\nu^\prime$, and one bosonic Matsubara frequency $\omega$ one can define two convenient frequency conventions: the particle-hole (ph) $\chi^{\nu,\nu^\prime\! ,\,\omega}$ and the particle-particle (pp) $\chi^{\nu,\nu^\prime\! ,\,\omega}_\text{pp}$ notation

\begin{subequations}\label{eq:phconv}
\begin{align}
    \textbf{ph: } \nu_1&=\nu & \textbf{pp: }\nu_1&=\nu\\
    \nu_2&=\nu+\omega & \nu_2&=\omega-\nu^\prime \\
    \nu_3&=\nu^\prime+\omega & \nu_3&=\omega-\nu \\
    (\nu_4&=\nu^\prime) & (\nu_4&=\nu^\prime)
\end{align}
\end{subequations}

One can go from ph to pp notation by shifting $\omega\rightarrow \omega-\nu-\nu^\prime$.

Instead of using the definitions in \mbox{\cref{eq:gensus,eq:gensus_bar}} directly, it is often more useful to consider specific spin combinations, which correspond to physical response functions.

The spin combinations (channels) that we consider are charge (c), longitudinal spin (s), the coupling between charge and longitudinal spin (cs) and (sc), transversal spin ($S_x=S_y$) and the pairing channel (pair):

\begin{align}
\label{eq:channels_begin}
    \chi_\text{c}^{\nu,\nu^\prime\! ,\,\omega}&=\chi_{\uparrow\uparrow}^{ \nu,\nu^\prime\! ,\,\omega}+\chi_{\downarrow\downarrow}^{ \nu,\nu^\prime\! ,\,\omega}+\chi_{\uparrow\downarrow }^{ \nu,\nu^\prime\! ,\,\omega}+\chi_{\downarrow\uparrow, }^{ \nu,\nu^\prime\! ,\,\omega},\\
\label{eq:channels_s}
    \chi_\text{s}^{ \nu,\nu^\prime\! ,\,\omega}&=\chi_{\uparrow\uparrow }^{ \nu,\nu^\prime\! ,\,\omega}+\chi_{\downarrow\downarrow }^{ \nu,\nu^\prime\! ,\,\omega}-\chi_{\uparrow\downarrow }^{ \nu,\nu^\prime\! ,\,\omega}-\chi_{\downarrow\uparrow }^{ \nu,\nu^\prime\! ,\,\omega},\\
\label{eq:channels_cs}
    \chi_\text{cs}^{ \nu,\nu^\prime\! ,\,\omega}&=\chi_{\uparrow\uparrow }^{ \nu,\nu^\prime\! ,\,\omega}-\chi_{\downarrow\downarrow }^{ \nu,\nu^\prime\! ,\,\omega}-\chi_{\uparrow\downarrow }^{ \nu,\nu^\prime\! ,\,\omega}+\chi_{\downarrow\uparrow }^{ \nu,\nu^\prime\! ,\,\omega},\\
\label{eq:channels_sc}
    \chi_\text{sc}^{ \nu,\nu^\prime\! ,\,\omega}&=\chi_{\uparrow\uparrow }^{ \nu,\nu^\prime\! ,\,\omega}-\chi_{\downarrow\downarrow }^{ \nu,\nu^\prime\! ,\,\omega}+\chi_{\uparrow\downarrow }^{ \nu,\nu^\prime\! ,\,\omega}-\chi_{\downarrow\uparrow }^{ \nu,\nu^\prime\! ,\,\omega},\\
\label{eq:channels_sx}
    \chi_{S_x}^{\nu,\nu^\prime\! ,\,\omega}&=\chi_{S_y}^{\nu,\nu^\prime\! ,\,\omega}= \chi_{\overline{\uparrow\downarrow}}^{\nu,\nu^\prime\! ,\,\omega}+\chi_{\overline{\downarrow\uparrow}}^{\nu,\nu^\prime\! ,\,\omega},\\
\label{eq:channels_end}
    \chi_{\text{pair}}^{\nu,\nu^\prime\! ,\,\omega} &= -\chi_{\overline{\uparrow\downarrow},\text{pp}}^{\nu,\nu^\prime\! ,\,\omega} -\Big(\chi_{\overline{\downarrow\uparrow},\text{pp}}^{\nu,\nu^\prime\! ,\,\omega}\Big)^*.
\end{align}

For the subsequent discussion it is convenient to view the generalized susceptibilities as (infinite dimensional) matrices in the frequencies $\nu$, $\nu'$  and treat $\omega$ as an additional parameter. 

To lighten the notation we are going to omit the fermionic frequencies $\nu$ and $\nu^\prime$ in the following and state the bosonic frequency $\omega$ only explicitly, when it is not the same for all quantities in an expression.  Within this notation all operations (e.g. multiplication, transposition, inversion, ...) are meant as matrix operations in the fermionic Matsubara frequency space.

Note that for broken SU(2)-symmetry the previously separate charge and the longitudinal spin channels get coupled into the longitudinal (L) channel 

\begin{align}
\label{eq:chi_full}
\chi_\text{L} =     
\begin{pmatrix}
\chi_\text{c} & \chi_\text{cs}\\
\chi_\text{sc} & \chi_\text{s}
\end{pmatrix},
\end{align}
where the space of the longitudinal channel is a tensor product space of the charge and spin  fermionic Matsubara frequency spaces.

The generalized susceptibility matrices show interesting properties that restrict their eigenvalues and eigenvector structure and depend on the symmetries of the considered model (see \mbox{\cref{app:sus_syms}} for an in depth discussion).

For $\omega=0$, the matrices of the transversal spin channel $\chi_{S_x}$ as well as the pairing channel $ \chi_{\text{pair}}$ can be identified as centro-Hermitian matrices~\cite{reitner2023} (see \cref{eq:centro_hermitian}), and therefore only have eigenvalues that are real or come in complex conjugate pairs \cite{lee1980}. 
Moreover, the matrix of the coupled longitudinal channel $\chi_\text{L}$ is classified as a $\kappa-$real matrix \cref{eq:kappa_real}, which shows the same eigenvalue properties \cite{hill1992}. 

\begin{figure}[t!]
\includegraphics[scale=0.4]{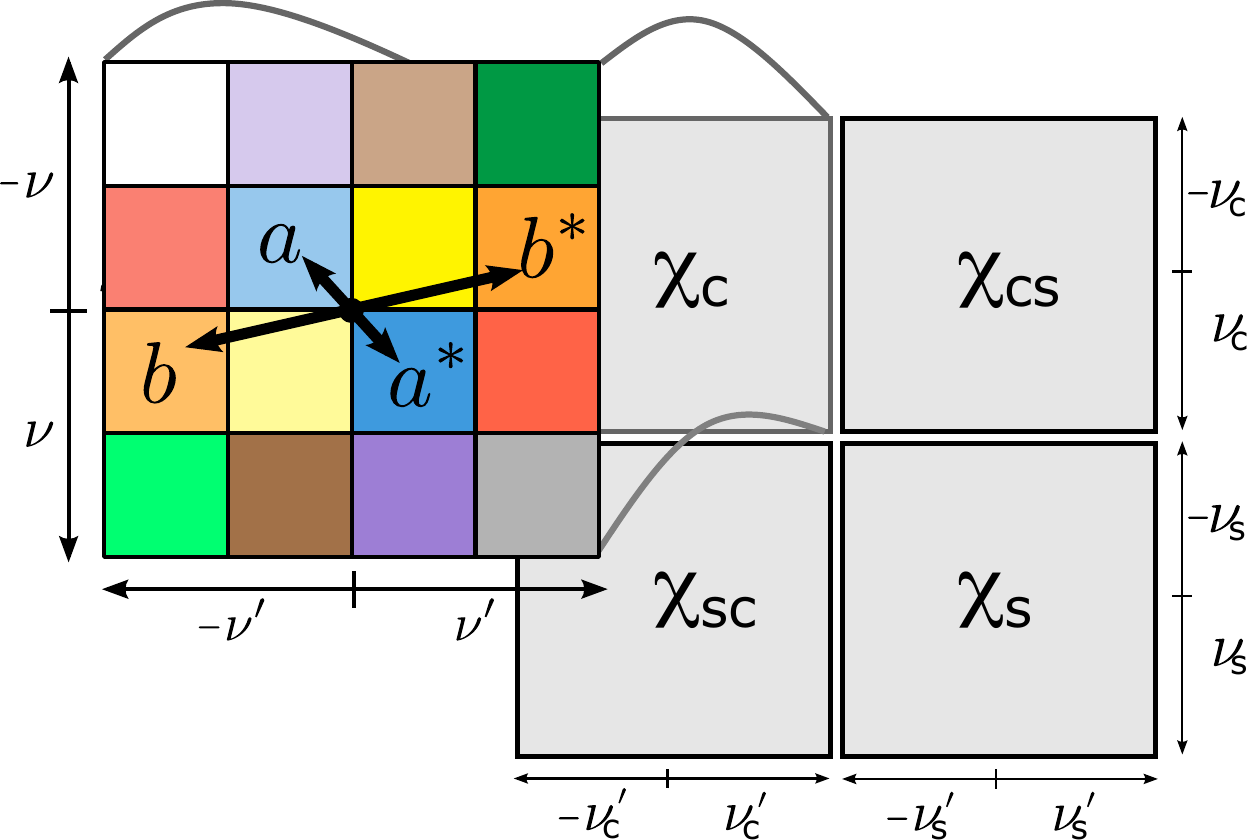}
\caption{\label{fig:matrix} Schematic illustration of a $\kappa-$real matrix that is built from blocks of centro-Hermitian matrices, which is relevant for the coupled longitudinal channel if $SU(2)_S$-symmetry is violated.}
\end{figure}

In addition, the generalized susceptibilities for all considered channels are symmetric matrices and, hence, can be diagonalized by an orthogonal transformation \cite{craven1969}. 
This, together with the centro-Hermitian respectively $\kappa-$real property, leads to the fact that eigenvectors $v_\alpha$ corresponding to real eigenvalues $\lambda_\alpha$ have a symmetric/antisymmetric real part  $\Re v_\alpha$ (in $\nu$) and an antisymmetric/symmetric imaginary part $\Im v_\alpha$ (see \cref{eq:centro_hermitian_ev}) ~\cite{reitner2023}. For the longitudinal channel eigenvectors are symmetric/antisymmetric in the charge respectively spin subspace (see \cref{eq:kappa_real_ev,eq:kappa_real_ev}). 
\cref{fig:matrix} shows a schematic example of a $\kappa-$real matrix that is built from blocks of centro-Hermitian matrices in the fermionic Matsubara frequencies, as it is the case in the coupled longitudinal channel.

A concise summary of the generalized susceptibility matrix properties for $\omega=0$ in the different channels and under additional symmetries can be seen in \cref{tab:symtable}, where \emph{No Symmetry} refers to time translation symmetry, spin conservation and that the Hamiltonian is a real function of $c$ and $c^\dagger$. Note that the last assumption can be  violated, for example, by the inclusion of an external electromagnetic potential via the Peierls substitution in lattice models and/or if spin-orbit coupling in is taken into account. All these symmetries can be derived by applying \cref{eq:chi_sym_cc,eq:chi_sym_cs,eq:chi_sym_HinR,eq:chi_sym_su2,eq:chi_sym_su2p} on the respective definition of the different channels.

\begin{table*}[t!]
\begin{tabular}{|c|c|c|c|}
\hline
Symmetry    & $\chi_\text{L}^{\omega=0}$                                  & $\chi_{Sx}^{\omega=0}$                       & $\chi_\text{pair}^{\omega=0}$                     \\ \hline
No symmetry & $\kappa$-real, symmetric                 & centro-Hermitian, symmetric & centro-Hermitian, symmetric \\ \hline
$SU(2)_S$      & block-diagonal, $\kappa$-real, symmetric & centro-Hermitian, symmetric & bi-symmetric, real          \\ \hline
$SU(2)_P$     & $\kappa$-real, symmetric                 & bi-symmetric, real          & centro-Hermitian, symmetric \\ \hline
PH ($SU(2)_S \bigotimes SU(2)_P$)      & block-diagonal, real, symmetric       & bi-symmetric, real          & bi-symmetric, real          \\ \hline
\end{tabular}
\caption{Matrix properties of the generalized susceptibilities under additional symmetries where \emph{No Symmetry} refers to  time translation symmetry, spin conservation and that the Hamiltonian is a real function of $c$ and $c^\dagger$.\\
For the additional symmetries we use: $SU(2)_S$ for Spin SU(2) symmetry $(h=0)$,  $SU(2)_P$ for Pseudo-Spin $SU(2)$ symmetry $(\delta\mu=0)$, PH for particle-hole symmetry ($SU(2)_S$ and $SU(2)_P$ symmetry).}
\label{tab:symtable}
\end{table*}

\subsection{Shiba Transformation}

The mapping---we apply in this paper---systematically
maps all quantities of a model with on-site interaction onto a corresponding model with sign-flipped interaction.  It has been used for the study in Ref.~\cite{springer2020} and is derived by applying a partial (local) particle-hole transformation, or Shiba transformation \cite{shiba1972}, acting on the creation and annihilation operators with

\begin{align}
\label{eq:shiba}
    c_\uparrow\rightarrow c_\uparrow, c^\dagger_\uparrow\rightarrow c^\dagger_\uparrow, c_\downarrow \rightarrow c^\dagger_\downarrow \text{ and } c^\dagger_\downarrow \rightarrow c_\downarrow.
\end{align}

By considering a local single orbital Hubbard Hamiltonian (written in a symmetric form with $\delta\mu = \mu -U/2$) one readily sees that applying the Shiba transformation

\begin{gather}
\begin{split}
\label{eq:shibaH}
    H(U,&\delta\mu,h) = \\
    - \delta\mu(n_\uparrow+n_\downarrow) - h(n_\uparrow-n_\downarrow) &+ U(n_\uparrow-1/2)(n_\downarrow-1/2)\\
    &\Bigg\updownarrow \text{Shiba} \\
    - \delta\mu(n_\uparrow-n_\downarrow) - h(n_\uparrow+n_\downarrow) &- U(n_\uparrow-1/2)(n_\downarrow-1/2) \\
    = H(-&U,h,\delta\mu),
\end{split}
\end{gather}
($n_\sigma=c^\dagger_\sigma c_\sigma$) corresponds to interchanging the values of $\delta\mu \leftrightarrow h$ as well as a sign flip in the interaction $U$ (any constant terms that appear from the mapping in the Hamiltonian have been omitted).

The Hamiltonian in \mbox{\cref{eq:shibaH}}  corresponds to the Hubbard atom (HA), which we consider for our exemplary study in \mbox{\cref{sec:application}} for different fillings and external magnetic fields $h$. The model can be solved analytically~\cite{pairault2000}; the corresponding expressions for the one- and two-particle Green's functions can be found in \mbox{\cref{app:sus_HA}}

The (non-local) Shiba mapping can be also applied for a bipartite lattice model, i.e. a lattice that can be divided into two sub-lattices, where the hopping from each site only connects to the other sub-lattice. The Hamiltonian of such a Hubbard model (HM) reads

\begin{align}
\label{eq:H_bipartite}
\begin{split}
        H &= -\sum_{i,j,\sigma}t_{i,j}c^\dagger_{i,\sigma}c_{j,\sigma} - \delta\mu\sum_i(n_{i,\uparrow}+n_{i,\downarrow}) \\
        &- h\sum_i(n_{i,\uparrow}-n_{i,\downarrow}) + U\sum_i(n_{i,\uparrow}-1/2)(n_{i,\downarrow}-1/2),
\end{split}
\end{align}
where $t_{ij}$ is zero if $i$ and $j$ belong to the same sub-lattice. Assigning one sub-lattices to all even and the other one to all odd sites respectively we readily see that the non-local Shiba transformation $c_{i,\downarrow}\rightarrow c^\dagger_{i,\downarrow}(-1)^i$ leaves the hopping term \mbox{\cref{eq:H_bipartite}} invariant and otherwise follows \mbox{\cref{eq:shibaH}}. Therefore, the effect of the Shiba transformation on bipartite systems (with local on-site interaction) can be regarded to be analog to purely local systems. 

The Shiba mapping is linked to two sets of $SU(2)$ spin algebras, of which the generators are the (local) spin operator $\boldsymbol{\mathcal{S}}$ with the components 

\begin{align}
    \mathcal{S}_x &= c_\uparrow^\dagger c_\downarrow + c_\downarrow^\dagger c_\uparrow,\\
    \mathcal{S}_y &= -i(c_\uparrow^\dagger c_\downarrow - c_\downarrow^\dagger c_\uparrow),\\
    \mathcal{S}_z &= c_\uparrow^\dagger c_\uparrow - c_\downarrow^\dagger c_\downarrow
\end{align}
and the (local) pseudospin operator $\boldsymbol{\mathcal{S}}_p$ with the components

\begin{align}
    \mathcal{S}_{p,x} &= c_\uparrow^\dagger c_\downarrow^\dagger + c_\downarrow c_\uparrow,\\
    \mathcal{S}_{p,y} &= -i(c_\uparrow^\dagger c_\downarrow^\dagger - c_\downarrow c_\uparrow),\\
    \mathcal{S}_{p,z} &= c_\uparrow^\dagger c_\uparrow + c_\downarrow^\dagger c_\downarrow -1.
\end{align}

Evidently, the Shiba transformation maps the spin $\boldsymbol{\mathcal{S}}$ to the pseudospin $\boldsymbol{\mathcal{S}}_p$. Since the components of $\boldsymbol{\mathcal{S}}$ and $\boldsymbol{\mathcal{S}}_p$ commute, spin and pseudospin can be regarded as independent
set of operators\cite{rohringer2013}.

The pseudospin can be given the following physical interpretation: $\mathcal{S}_{p,x}$ and $\mathcal{S}_{p,y}$ correspond to the local Cooper pair operators, which describe the real and imaginary part of the order parameter of a strong-coupling (Bose-Einstein-like) superconductor, and $\mathcal{S}_{p,z}$ describes the deviation of the density from half filling. 

The Hamiltonian in the upper line of in \cref{eq:shibaH} commutes with all components of $\boldsymbol{\mathcal{S}}$ if $h=0$, which we will denote  as $SU(2)_S$ symmetry, and with all components of $\boldsymbol{\mathcal{S}}_p$ for $\delta\mu=0$, referred to as $SU(2)_P$ symmetry.

Systems that exhibit a symmetry under rotations generated by spin \emph{and} pseudospin, are symmetric under $SO(4)\simeq (SU(2)_S \otimes SU(2)_P)/\mathbb{Z}_2$ \cite{rohringer2013}. These $SO(4)$ symmetric systems are particle-hole symmetric\footnote{Note that, for the scope of this paper, we adopt the standard definition of particle-hole transformation, i.e. $c_\sigma\rightarrow c^\dagger_\sigma$. Of course, in principle, one could also choose the alternative definition of $c_\sigma\rightarrow c^\dagger_{-\sigma}$. The latter definition would be consistent with the fact that the particle-hole transformation changes the charge of the particle and therefore also its gyro-magnetic moment. This would then be accounted for by flipping the spin in the transformation. Using this alternative definition, the $SU(2)_P$ symmetry would become equivalent to the particle-hole symmetry and the $SO(4)$ symmetry would be achieved in the presence of simultaneous particle-hole and $SU(2)_S$ symmetry.}.

The on-site physical susceptibilities, describing the (linear) response of the system on a local observable $A$ when a local observable $B$ is coupled to an external perturbation can be calculated by Fourier-transforming/analytically continuing the following expression:
\begin{align}
    \chi_\text{AB}^\text{phys} (\tau) = \langle A(\tau) B(0)\rangle - \langle A\rangle\langle B\rangle. 
\end{align}

In Matsubara frequency space, they are related to the generalized susceptibilities via:
\begin{align}
\label{eq:chi_phys}
    \chi_r^{\text{phys}}(\omega)=\frac{1}{\beta^2}\sum_{\nu\nu^\prime}\chi_{r}^{\nu\nu^\prime\omega}.
\end{align}

Using \cref{eq:chi_phys} and the definitions in \cref{eq:channels_begin,eq:channels_s,eq:channels_cs,eq:channels_sc,eq:channels_sx,eq:channels_end} we find

\begin{align}
    \chi_\text{c}^\text{phys}&: A=B=\mathcal{S}_{p,z},\\
    \chi_\text{s}^\text{phys}&: A=B = \mathcal{S}_z,\\
    \chi_\text{cs}^\text{phys}&: A=\mathcal{S}_{p,z}, B=\mathcal{S}_z,\\
    \chi_\text{sc}^\text{phys}&: A=\mathcal{S}_z, B=\mathcal{S}_{p,z},\\
    \chi_{S_x}^\text{phys}&: A=B=\mathcal{S}_x,\\
    \chi_{\text{pair}}^\text{phys}&: A=B=\mathcal{S}_{p,x}.
\end{align}

The mapping of the physical susceptibilities under the Shiba transformation is thereby clearly identified from the corresponding mapping of the spin and pseudospin components.

The more complicated case of the Shiba mapping for the local generalized susceptibilities is derived in the next \cref{sec:der_mapp}.

\section{Derivation of the mapping}
\label{sec:der_mapp}

In the following, we derive how the Shiba transformation in \cref{eq:shiba} is acting on the local\footnote{Note that all derivations in this section also hold for non-local generalized susceptibilities of bipartite lattice models if the following criterion is fulfilled: the algebraic sum of the lattice indices of the considered generalized susceptibility is even.} generalized susceptibilities of the charge, longitudinal spin, transversal spin, and pairing channel (see \cref{eq:channels_begin,eq:channels_s,eq:channels_cs,eq:channels_sc,eq:channels_sx,eq:channels_end}) for the most general cases, i.e. arbitrary filling, external magnetic field, and arbitrary bosonic transfer frequency $\omega$. Thereby, the coupling of
the charge and longitudinal spin channel for broken $SU(2)_S$ symmetry, i.e. finite magnetic field $h$, into the longitudinal channel (c.f. \cref{eq:chi_full}) has to be explicitly considered. 

By applying the Shiba transformation on the general definition in \mbox{\cref{eq:gensus}} we find the relations:

\begin{align}
\label{eq:chi_shiba}
\begin{split}
    \chi_{\uparrow\uparrow,\text{U}}^{\nu,\nu^\prime\! ,\,\omega} = \chi_{\uparrow\uparrow,\text{-U}}^{\nu,\nu^\prime\! ,\,\omega}, \; \;\chi_{\downarrow\downarrow,\text{U}}^{\nu,\nu^\prime\! ,\,\omega} = \chi_{\downarrow\downarrow,\text{-U}}^{-\nu-\omega,-\nu^\prime-\omega ,\,\omega}, \\ \chi_{\uparrow\downarrow,\text{U}}^{\nu,\nu^\prime\! ,\,\omega} = -\chi_{\uparrow\downarrow,\text{-U}}^{\nu,-\nu^\prime-\omega,\,\omega}, \; \;\text{and}\;\;\chi_{\downarrow\uparrow,\text{U}}^{\nu,\nu^\prime\! ,\,\omega} = -\chi_{\downarrow\uparrow,\text{-U}}^{-\nu-\omega,\nu^\prime\! ,\,\omega},
\end{split}
\end{align}
where the subscript U refers to the model parameters $(U,\delta\mu,h)$ and the subscript -U to $(-U,h,\delta\mu$), i.e. the two models related by the Shiba transformation in \cref{eq:shibaH}.

With these building blocks we can now apply the Shiba mapping to the coupled longitudinal channel.

We start with the case of $\omega=0$ and then generalize to finite bosonic frequencies. 

We will identify the similarity transformation $S$ that captures the action of the Shiba transformation onto the generalized susceptibilities $\chi_\text{L,U}$ and $\chi_\text{L,-U}$. Thereby, we employ the following matrix $Q$ in Matsubara space:

\begin{align}
\label{eq:q_trafo}
    Q = \frac{1}{\sqrt{2}}    
    \begin{pmatrix}
    \mathbb{1} & -J\\
    \mathbb{1} & J
    \end{pmatrix}\quad \text{with} \quad J=\mqty(\admat{1,\udots,1}),
\end{align}
where we divided the Matsubara frequencies into quadrants of positive $+$ and negative $-$ frequencies
\begin{equation}
\label{eq:quadrants}
    \left( \begin{array}{c|c}
   \chi^{--} & \chi^{-+} \\
   \hline
   \chi^{+-} & \chi^{++} \\
\end{array}\right).
\end{equation}

The similarity transformation $Q^T \chi_{CS} Q$ separates the symmetric and anti-symmetric part of a centro-symmetric matrix $(JAJ=A)$ into  two blocks along the diagonal and the symmetric and anti-symmetric part of a centro-skew-symmetric matrix $(JAJ=-A)$ into two blocks along the skew diagonal.

Hence, a similarity transformation with $Q$ separates the real $\chi^\prime$ and imaginary part $\chi^{\prime\prime}$ of  a centro-Hermitian matrix  $\chi_\text{CH}$ $(J\chi_\text{CH}J=\chi_\text{CH}^*)$ into the symmetric and anti-symmetric components respectively

\begin{align}
\label{eq:q_trafo_ch}
    Q\chi_\text{CH} Q^T =  
    \begin{pmatrix}
        \chi_\mathcal{A}^\prime & i\chi_\mathcal{S}^{\prime\prime}\\
        i\chi_\mathcal{A}^{\prime\prime} & \chi_\mathcal{S}^{\prime}
    \end{pmatrix}.
\end{align}

A detailed definition of the different blocks can be found in \mbox{\cref{app:mapping}}.

By defining the transformations $\mathcal{Q}$ and $T$ in the tensor product space of the coupled L channel

\begin{align}
\label{eq:QandT_trafo}
    \mathcal{Q}= 
    \begin{pmatrix}
        Q & \mathbb{0}\\
        \mathbb{0} & Q
    \end{pmatrix} \quad \text{and} \quad T=
\left(
\begin{array}{cc}
    \begin{array}{c|c}
        \mathbb{1} & \mathbb{0} \\
        \hline
        \mathbb{0} & \mathbb{0}
    \end{array}
    &
    \begin{array}{c|c}
        \mathbb{0} & \mathbb{0} \\
        \hline
        \mathbb{0} & \mathbb{1}
    \end{array} \\
    \arrayrulecolor{white}\midrule
    \arrayrulecolor{black}
    \begin{array}{c|c}
        \mathbb{0} & \mathbb{0} \\
        \hline
        \mathbb{0} & \mathbb{1}
    \end{array}
    &
    \begin{array}{c|c}
        \mathbb{1} & \mathbb{0} \\
        \hline
        \mathbb{0} & \mathbb{0}
    \end{array}
\end{array}
\right)
\end{align}
(c.f. \mbox{\cref{eq:chi_full}}) we can determine (see \mbox{\cref{app:mapping}}) the transformation $S$, which relates the two sides of the Shiba transformation: 

\begin{align}
\label{eq:chi_shiba_main}
    \chi_\text{L,U} = S \chi_\text{L,-U} S \quad \text{with} \quad S = \mathcal{Q}^T T \mathcal{Q}.
\end{align}

In \mbox{\cref{app:mapping}} we show that this relation holds for arbitrary bosonic transfer frequencies $\omega$, when adapting the $Q$ and $T$ matrices by shifting the boundaries of the Matsubara frequency quadrants  by $(-\omega/2,-\omega/2)$ in analogy to \mbox{\cref{eq:quadrants}}. 

Hence, for finite $\omega$ the susceptibility matrices can be viewed as centro-Hermitian in regard to the shifted Matsubara frequency basis (see \cref{eq:freq_shift}).

From \mbox{\cref{eq:chi_shiba_main}} (better visible in \cref{eq:chi_full_shiba}) it follows that the Shiba transformation leaves the real anti-symmetric part $\chi^\prime_\mathcal{A}$ of each charge-spin components ($\chi_c,\chi_s,\chi_{cs},\chi_{sc}$) in $\chi_\text{L}$  invariant and exchanges the real symmetric part $\chi^\prime_\mathcal{S}$ between $\chi_\text{c} \leftrightarrow \chi_\text{s}$, and between $\chi_\text{cs} \leftrightarrow \chi_\text{sc}$

\begin{align}
\label{eq:mapping_rules_re}
\begin{split}
    \chi_{\text{c,-U},\mathcal{A}}^\prime &\longleftrightarrow \chi_{\text{c,U},\mathcal{A}}^\prime,\\
    \chi_{\text{s,-U},\mathcal{A}}^\prime &\longleftrightarrow \chi_{\text{s,U},\mathcal{A}}^\prime,\\
    \chi_{\text{c,-U},\mathcal{S}}^\prime &\longleftrightarrow \chi_{\text{s,U},\mathcal{S}}^\prime,\\
    \chi_{\text{cs,-U},\mathcal{A}}^\prime &\longleftrightarrow \chi_{\text{cs,U},\mathcal{A}}^\prime,\\
    \chi_{\text{sc,-U},\mathcal{A}}^\prime &\longleftrightarrow \chi_{\text{sc,U},\mathcal{A}}^\prime,\\
    \chi_{\text{cs,-U},\mathcal{S}}^\prime &\longleftrightarrow \chi_{\text{sc,U},\mathcal{S}}^\prime.
\end{split}
\end{align}
For the imaginary parts neither the symmetric nor the anti-symmetric part is invariant under the mapping and they exchange according to the rules below:
\begin{align}
\label{eq:mapping_rules_im}
\begin{split}
    \chi_{\text{cs,-U},\mathcal{A}}^{\prime\prime} &\longleftrightarrow \chi_{\text{s,U},\mathcal{A}}^{\prime\prime},\\
    \chi_{\text{sc,-U},\mathcal{A}}^{\prime\prime} &\longleftrightarrow \chi_{\text{c,U},\mathcal{A}}^{\prime\prime},\\
    \chi_{\text{c,-U},\mathcal{S}}^{\prime\prime} &\longleftrightarrow \chi_{\text{cs,U},\mathcal{S}}^{\prime\prime},\\
    \chi_{\text{sc,-U},\mathcal{S}}^{\prime\prime} &\longleftrightarrow \chi_{\text{s,U},\mathcal{S}}^{\prime\prime}.
\end{split}
\end{align}

By examining the eigenvectors of the generalized susceptibility in the longitudinal channel, it follows that the symmetric part in one subspace is mapped from the charge to the spin subspace, and vice versa. However, the antisymmetric part remains unaffected by this transformation (refer to \cref{eq:map_ev_even,eq:map_ev_odd}).

Finally, we apply the Shiba transformation on the generalized susceptibilities of the pairing and the transversal spin channel. 
From the transformation of \cref{eq:gensus_bar} we get

\begin{align}
\label{eq:map_bar}
    \chi_{S_x,\text{U}}^{\omega}=\chi_\text{pair,-U}^{-\omega}
\end{align}
(details in \cref{app:mapping}).

\section{Application of the mapping}
\label{sec:application}

The mapping derived in \cref{sec:der_mapp} can be useful in several practical contexts, being applicable also beyond the purely local Hubbard atom Hamiltonian in the first line of \cref{eq:shibaH}. 
In fact, the expressions derived in the previous section may also be applied to the local two-particle quantities of bipartite lattice models with on-site electrostatic interaction, where their mapping  would work in a  completely analogous fashion (see \cref{eq:H_bipartite}). For instance, our equations could be directly exploited to map the on-site generalized susceptibilities of the Hubbard model on a generic $d-$dimensional bipartite lattice with nearest neighbor\footnote{We recall that the hopping $t_{ij}$ transforms under the Shiba transformation for lattice models as: $t_{ij} \rightarrow t_{ij}$ if one index belongs to an odd site and the other on an even site and $t_{ij} \rightarrow \sigma t_{ij}$ otherwise. Hence, if the hopping connects odd to odd or even to even sides (e.g.~ via a n.n.n hopping $t^\prime$), this hopping term would change under the Shiba mapping from $t^\prime$ to $-t^\prime$ for the down electrons only, limiting the potential usefulness of the mapped expressions.} hopping $t$ and $(U,\delta\mu,h)$ to the corresponding lattice model with the same hopping and $(-U,h,\delta\mu)$~\footnote{On an even more general perspective, it is worth mentioning that \cref{eq:chi_shiba_main,eq:map_bar} would hold for the on-site generalized susceptibilities of arbitrary lattice models, albeit with the important caveat that the specific meaning of subindices U and -U need to be properly reinterpreted. In practice, this can be done by determining the precise effect of the Shiba transformation on the Hamiltonian of the considered model and then defining the meaning of the labels U and -U according to this result.}.
Evidently, this feature renders the derived mapping also immediately applicable to the local two-particle quantities of the widely used Dynamical Mean-Field Theory (DMFT) \cite{georges1996}.

Independent of the specific models considered, an evident advantage of the mapping is to potentially reduce the effort of two-particle calculations, if these need to be systematically performed in different regions of the phase diagram. Further, one could exploit the mapping to investigate the two-particle properties of regimes (e.g., with $SU(2)_S$-broken symmetry) that might otherwise not be directly and/or easily accessible with the numerical algorithms at disposal.

Finally, the derived mapping gives also some fundamental insights on how two-particle quantities of different scattering channels are connected to each other and how they are linked to the symmetries of the system.

As an example to illustrate how this aspect of the mapping works, we look at the problem of the two-particle irreducible vertex divergences in the Hubbard atom, which is of considerable interest in the context for the investigation of the breakdown of the self-consistent perturbation expansion for many-electron systems.

\subsection{Vertex divergences and their generalization}
\label{sec:vertexdivs}

In this section, we extend the existing studies\cite{schafer2016,gunnarsson2017,thunstrom2018,springer2020} on the divergences in the two-particle irreducible vertex of the Hubbard atom, which have been hitherto restricted to the rather specific, highly symmetric case of $\delta \mu= h= 0$ (i.e., of perfect particle-hole symmetry, so $SU(2)_S$ and $SU(2)_P$ symmetry).
In particular, we aim here at individuating the divergences of the two-particle irreducible vertices at transfer frequency $\omega=0$ in all scattering channels for the {\sl whole} parameter space of the Hubbard atom with positive (repulsive) as well as negative (attractive) on-site interaction, arbitrary filling, and in the presence of an external magnetic field $h$.

We  briefly recall here that the two-particle irreducible vertex function $\Gamma_r^{\nu,\nu'}(\omega)$ in a given scattering channel $r$ (e.g., longitudinal, pairing, etc.) is formally defined\cite{abrikosov1975,bickers2004,rohringer2012,rohringer2013} as the kernel of the (self-consistent) Bethe-Salpeter equation (BSE) in the corresponding sector $r$, and, as such, can be computed by inverting the BSE, i.e. $$\Gamma_r^{\nu\nu'}(\omega) \! := \! \left[\chi_r^{-1}\right]^{\nu,\nu^\prime\! ,\,\omega} - \left[\chi_{r,0}^{-1}\right]^{\nu,\nu^\prime\! ,\,\omega},$$ 

where the transversal channels has to be treated with care if $H\notin \mathbb{R}$ and $\omega\neq 0$ (see \cref{app:BSE_transversal})

Hence, the divergences of the $\Gamma_r$ are directly determined by the vanishing of an eigenvalue of the corresponding generalized susceptibility matrix $\chi_r^{\nu,\nu^\prime\! ,\,\omega}$ in the fermionic Matsubara frequencies $\nu,\nu^\prime$~\footnote{None of the different ``bubble'' terms $\chi_{r,0}^{\nu,\nu^\prime\! ,\,\omega}$ can trigger a divergences, as they are just diagonal in frequency spaces and defined by the product of two Green's functions.}.

As mentioned in \cref{sec:gen_sus} by perfect $SU(2)_S$ and $SU(2)_P$ symmetry, all local two-particle quantities for zero-transfer frequency ($\omega=0$) are real and symmetric matrices. Hence, the \emph{sign change} of an eigenvalue can only happen if this eigenvalue vanishes. This eigenvalue is then responsible for the divergence of $\Gamma_r^{\nu\nu^\prime}$. These sign changes, as discussed in the recent literature \cite{gunnarsson2017,springer2020,vanloon2020,reitner2020,chalupa2021,adler2024}, are crucial to drive the suppression/enhancements of the corresponding physical static response of the systems, and, in more complicated lattice systems, even to trigger thermodynamic phase-instabilities \cite{nourafkan2019,vanloon2020,reitner2020,reitner2023,kowalski2023,vanloon2024}.
However, if either $SU(2)_S$ ($h \neq 0$) or $SU(2)_P$ ($\delta \mu \neq 0$) symmetries are broken, $\chi_{r}^{\nu\nu^\prime}$  has no longer to be a real and symmetric matrix for all channels, but can be either a centro-Hermitian or a $\kappa$-real matrix (see \cref{tab:symtable}).
Consequently, its eigenvalues are no longer required to be real: They can be either real or appear in complex conjugate pairs, consistent with the fundamental theorem of algebra \cite{lee1980,hill1992}.
 
As a consequence, the discussed sign-changes of a eigenvalue of the generalized susceptibilities, or at least of their real part, may now occur continuously in the parameter space {\sl without} the necessity that the eigenvalue vanishes, which would trigger a vertex divergence. 
In fact, complex conjugate pairs of eigenvalues can switch the sign of their real parts without vanishing by keeping their imaginary parts finite. We note that the occurrence of this specific behavior of the eigenvalues of the generalized charge susceptibility has been reported in the phase diagram of Ref.~\cite{vucicevic2018} showing the DMFT\cite{georges1996}/Cellular-DMFT\cite{kotliar2001,maier2005} solution of the HM out of half-filling, as well as, in a somewhat different perspective, in Refs.~\cite{reitner2020,reitner2023,kowalski2023}. Indeed, these observations highlight the general relevance of such kind of evolution of the eigenvalues of the generalized susceptibility beyond the specific framework of the HA, which we are going to investigate systematically in the following.
In this context, we define the occurrence of a complex conjugate pair of eigenvalues with vanishing real part, but finite imaginary part as \emph{pseudo-divergence}.
In this way, the concept of vertex divergences can be extended to those cases, where sign-changes of the real part of an eigenvalue pair of $\chi_r^{\nu\nu'}$ are not directly associated to an actual divergence of $\Gamma_r^{\nu\nu'}$.
Loosely speaking, this generalization might recall the case of a branch cut of the square root function which is not crossing through the origin, but rather the negative real axis at a finite value by changing the sign of the imaginary part. 
Eventually, the pseudo-divergences defined here in the physical parameter space may be linked to the analytic properties of the generalized susceptibilities\cite{rohshap2023,rohshap2024} and/or of the perturbative expansion of the electron self-energy\cite{wu2017,rossi2016,vanhoucke2021,kim2022} in abstract complex planes obtained, e.g.~through the complexification of the electronic interaction $U$\cite{wu2017,rohshap2023,rohshap2024}.

\begin{figure*}[th!]
\includegraphics[width=\textwidth]{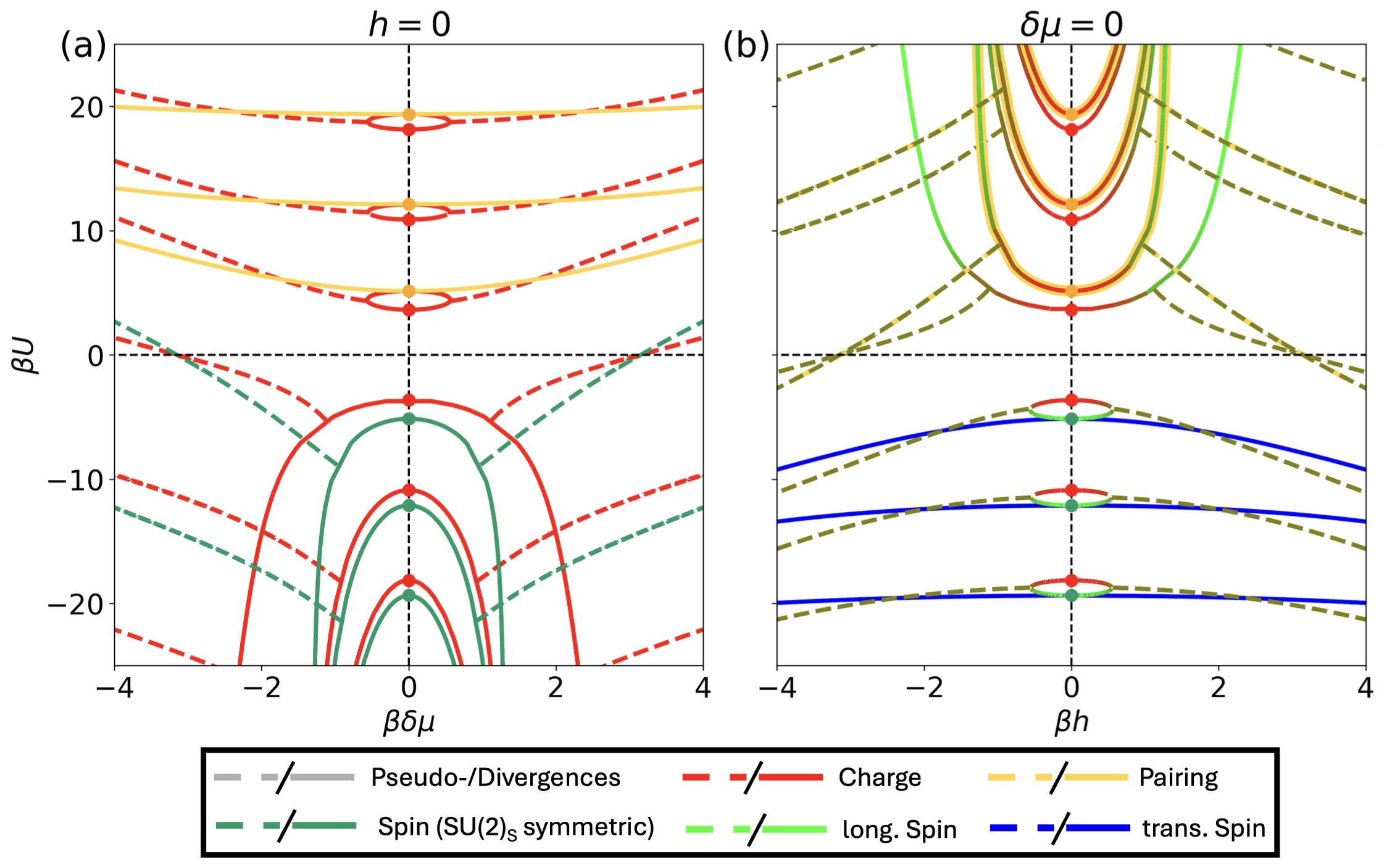}
\caption{\label{fig:divs_extreme} Dimensionless phase space diagram of the Hubbard atom with repulsive and attractive interaction, showing the location and the nature of the \mbox{(pseudo-)}divergences of the irreducible vertex functions in the different channels for the two special cases of \textbf{(a)} no magnetic field but out of half-filling (left panel) and \textbf{(b)} half filling in the presence of a magnetic field (right panel). Solid lines mark divergences and dashed lines mark pseudo-divergences, while, in order to identify the different channels, the following color coding is adopted (see legend in the box below the figure): \emph{red} for a vertex \mbox{(pseudo-)}divergence in the charge channel; \emph{yellow} for the pairing channel; \emph{green} for the longitudinal spin channel; \emph{blue} for the transversal spin channel; \emph{bluish green} for degenerated spin channel. Further, when $SU(2)_S$ symmetry is broken ($h \neq 0)$, the coupled longitudinal channel needs to be considered on a whole (see text): In this case, a weighted mixing of green and red has been used.}
\end{figure*}

\subsection{Dimensionless representation}

Before starting the analysis of the vertex-divergences and pseudo-divergences in the whole phase space of the Hubbard atom, it is convenient to make a general consideration based on the formal  Lehmann representation of the on-site\footnote{Note, however, that the presented arguments also apply for non-local susceptibilities with the only difference that the $\sigma$ index must be interpreted as combined $\sigma,k$ index} generalized susceptibilities, whose expression are reported, e.g., \cref{app:sus_HA} and in Ref.~\cite{toschi2007}.
To obtain a dimensionless representation we scale all parameters of the Hamiltonian with inverse temperature $\beta$. Formally this can be done by considering a model Hamiltonian, which is defined through $N$ real parameters  $\eval{\{\varepsilon_i\}}_{i=1}^N$ (e.g., the hopping $t$, the on-site interaction $U$, etc.), where all $\varepsilon_i$ have the dimension of an energy. Evidently, all eigenenergies of the many-electron Hamiltonian considered will only depend on these parameters (or combination thereof), i.e., $\{E_j\}=\{E_j(\{\varepsilon_i\})\}$. As a consequence, all corresponding dimensionless quantities $\beta E_j$ must only be dependent on the set $\{\beta\epsilon_i\}$ for all $j$. Then, by looking at the Lehmann representation of the generalized susceptibility, one readily realizes that, apart from an overall scaling of $\beta^3$, the generalized susceptibility can be always expressed in terms of the parameter set $\{\beta E_j\}$ which can be fully expressed via the parameter set $\{\beta \varepsilon_i\}$.
Therefore, one can rewrite the Lehmann representation as:
\begin{align}
\label{eq:red}
    \chi^{\nu\nu^\prime\omega}_{\alpha_1,\alpha_2,\alpha_3,\alpha_4} (\beta,\{\varepsilon_i\}) = \beta^3 f^{nn^\prime m}_{\alpha_1,\alpha_2,\alpha_3,\alpha_4}(\{\beta\varepsilon_i\}),
\end{align}
where $n,n^\prime$ and $m$ are the indices for $\nu,\nu^\prime$ and $\omega$ and $\alpha_i$ is a generic set of quantum numbers.
On the basis of this observation, presenting our results for the location and the properties of the vertex \mbox{(pseudo-)}divergences  of the HA in the \emph{dimensionless} parameter space $\{\beta\delta\mu, \beta h, \beta U\}$ turns out to be a particularly convenient choice, as it allows for a ``universal'' representation of our findings, directly applicable to any finite value of the temperature.

Before discussing the results, let us clarify the semantics.
The nature of the vertex \mbox{(pseudo-)}divergence is made clear by the color and the line-style in the plot: Solid lines mark divergences and dashed lines mark pseudo-divergences. 
For identifying the different channels, where the \mbox{(pseudo-)}divergence occur, we adopt the following color coding, extending the one introduced in existing literature:  We use \emph{red} for a vertex \mbox{(pseudo-)}divergence in the charge channel; \emph{yellow} for the pairing channel; \emph{green} for the longitudinal spin channel; \emph{blue} for the transversal spin channel; \emph{bluish green} for the spin channel at $SU(2)_S$ symmetry, where all three spin directions are equivalent. Eventually, for broken $SU(2)_S$ symmetry ($h \neq 0)$, where the coupled longitudinal channel needs to be considered on a whole, a weighted mixing of green and red is used. The proportion of red/green is determined by the percentage of the norm that the vector has in the charge/spin subspace for more information see \cref{app:weight}.
Further, we classify a divergence as ``symmetric/antisymmetric" if the corresponding eigenvector of the generalized susceptibility, which has a vanishing eigenvalue, has a symmetric/antisymmetric real part and an antisymmetric/symmetric imaginary part, provided that we use the norm \cref{eq:T_norm} to normalize the eigenvectors.

\subsection{Vertex divergences of the HA: Limiting cases}
\label{sec:limcas}

We start our analysis by illustrating in \cref{fig:divs_extreme} the results obtained for two selected planes in the dimensionless phase space introduced above. These correspond to the two special cases beyond perfect particle-hole/$SO(4)$ symmetry, where either the $SU(2)_S$ $(h=0$ \cref{fig:divs_extreme}a) or the $SU(2)_P$ $(\delta\mu=0$ \cref{fig:divs_extreme}b) symmetry is preserved. 
Such choice allows us to showcase in the most transparent way the action of the Shiba-mapping on the vertex functions beyond particle-hole symmetry. In fact, a first glance  at the geometrical shape and at the nature (encoded by the different colors) of the vertex (pseudo-)divergence
lines displayed in \cref{fig:divs_extreme} easily unveils the direct  connection between the results of the repulsive model for varying $\delta \mu$ (upper panel of \cref{fig:divs_extreme}a) and those of the attractive model for varying $h$ (lower panel of \cref{fig:divs_extreme}b). Consistent with the mapping relations, an analogous correspondence is also observed between our two-particle calculations for
 the attractive model at different $\delta \mu$ (lower panel of \cref{fig:divs_extreme}a) and those for the repulsive model at different $h$ (upper panel of \cref{fig:divs_extreme}b).

Before proceeding with the detailed analysis of the geometry and the nature of vertex (pseudo-)divergence lines in  \cref{fig:divs_extreme}, we recall that for the case of perfect particle-hole/$SO(4)$ symmetry ($ \delta \mu$ = $h =0$), it has been analytically demonstrated in the literature\cite{schafer2016,thunstrom2018,springer2020} that the HA displays a discrete set of vertex divergences, marked by colored dots in \cref{fig:divs_extreme}. In particular, for \mbox{$U>0$}, the red dots on the positive $\beta U$-axis of both panels of  \cref{fig:divs_extreme} indicate the location of the irreducible vertex divergences in the charge channel of the HA  associated to an anti-symmetric eigenvector, while the orange dots mark the simultaneous occurrence of  vertex divergences in the charge and in the pairing sector, associated to a symmetric eigenvector. Consistent with the mapping for perfect particle-hole symmetry \cite{springer2020}, one finds along the negative $\beta U$-axes of \cref{fig:divs_extreme}, exactly in mirrored positions w.r.t.~the $U>0$ case, alternating antisymmetric vertex divergences in the charge sector (red dots) and symmetric ones in the spin sector (bluish green dots).

We begin now our detailed discussion of the vertex divergences in the HA beyond particle-hole symmetry by examining the data shown in \cref{fig:divs_extreme}a, corresponding to the case of $h=0$ and varying $\delta \mu$. Here, it should be stressed that, as soon as the  $SU(2)_P$-symmetry is lifted for any $\delta \mu \neq 0$, the symmetric divergences in the charge and pairing channel, marked by the orange dots on the positive $\beta U$ axes of \cref{fig:divs_extreme}, do no longer occur simultaneously, yielding separate divergence lines (marked in yellow and red, respectively, for the pairing and the charge sector). However, this is not the only qualitative difference emerging between these two kind of vertex divergences for $\delta \mu \neq 0$: As reported in \cref{tab:symtable} the generalized susceptibility in the pairing channel is a real and bi-symmetric matrix as long as the $SU(2)_S$ symmetry is preserved. Therefore, differently from its charge counterpart, its eigenvalues remain purely real numbers and the associated vertex actually diverges, without any possible occurrence of pseudo-divergences and/or exceptional points (EPs). The latter are defined as the points in parameter space where two (or more) eigenvalues \emph{and} eigenvectors coalesce and the matrix is no longer diagonalizable~\cite{craven1969,reitner2023}. For the same reason, the symmetric nature of its singular eigenvectors is maintained for arbitrary values of $\delta\mu$.

On a more geometrical viewpoint, one can easily observe in \cref{fig:divs_extreme}a that the divergences lines of the charge sector form elliptic shaped loops in the phase space, terminating on both sides in EPs for positive and negative values of $\delta \mu$.
Henceforth, for the sake of conciseness, we will refer to these specific divergence structures, including the associated pseudo-divergences lines, with the short-hand term of \emph{loops}.
These loops each consist of one symmetric (upper loop part) and one antisymmetric (lower loop part) divergence line which meet at an EP. From this EP a pseudo-divergence line emerges.
Quantitatively, we note that the EPs of the loop found at the lowest $\beta U$ values corresponds to a hole/electron doping of about $6$\%, which roughly coincides~\footnote{While, at this stage, we might regard such a coincidence as purely accidental, we note that an almost identical doping level of about $6\%$ for the occurrence of the first EP has also been recently found in numerical calculations for the Anderson Impurity model in the wide-band limit\cite{LeonPA2022}.} with the doping level at which the corresponding EPs have been found in the DMFT/Cellular DMFT calculations of the HM of Ref.~\cite{vucicevic2018}.
It is also important to stress, that the three loops shown in \cref{fig:divs_extreme}, as all the other vertex-divergence structures discussed in the following, represent only the first three loop structures encountered by increasing $\beta U$. In fact, consistent with Refs.~\cite{schafer2016,thunstrom2018,pelz2023}, where the occurrence of infinitely many vertex divergences at half filling has been demonstrated, we  also find infinitely many loops for increasing $\beta U$ values.

Turning now to the study of the attractive sector, we find that the divergence lines show a completely different shape:
They are disposed in a series of parabolas one inside the other, as it can be seen on the bottom part of \cref{fig:divs_extreme}a. 
More specifically, considering first the charge sector, the geometrical maximum of the parabola is represented by the (antisymmetric) divergences located at particle-hole symmetry. Upon doping, an antisymmetric divergence line meets with a symmetric divergence line out of half filling at an EP, forming a pseudo-divergence line. The qualitatively different geometrical structure compared to the elliptical shaped divergence line structures in the repulsive sector corresponds to a specific difference: Each symmetric divergence line extends down to $\beta U\rightarrow-\infty$. The reason why this happens is the absence of symmetric divergences at half filling in the charge sector, which prevents the formation of closed loops of symmetric/antisymmetric divergences at difference with the repulsive sector. 
 
In the spin sector, we observe an analogous situation\footnote{Note that, also in this case, we only show the first three parabolas for each channel.}, only with interchanged roles of the eigenvector symmetries: The symmetric divergence lines, originating from the corresponding purely symmetric divergences at half-filling, meet with an antisymmetric divergence line, extended down to $\beta U\rightarrow-\infty$, at an EP, from which a pseudo-divergence emerges. 

A second important difference w.r.t.~the repulsive case is represented by the orientation of the pseudo-divergences of both channels. These, after emerging from the corresponding EPs, are directed towards {\sl smaller} absolute values of $\beta U$, in contrast to the pseudo-divergences in the repulsive sector. Even more remarkably, they do cross the $U=0$ axis, continuing afterwards in the repulsive sector. 
Hence, one may regard the pseudo-divergences lines that cross the $U=0$ axis as driven by the symmetry breaking fields ($\delta\mu$ and/or $h$) in contrast to the pseudo-divergences lines that do not cross the $U=0$ axis which would then be driven by the interaction $U$.
More specifically, one can classify the \mbox{(pseudo-)}divergences with an integer number $N$ identifying the positive Matsubara frequency at which their associated eigenvector displays its largest component. In this way, we find that the $N^{th}$ pseudo-divergence of both channels crosses the $U=0$ axis at $\beta\delta\mu=\beta\mu=\pm (2N-1)\pi$ (see \cref{app:vertex}). In fact, this has to be the same point, because the generalized susceptibilities of two-channels coincide for $U=0$, due to the absence of vertex corrections. For more detailed information see \cite{essl2023}.

As we noted at the beginning of this subsection, in the case of preserved $SU(2)_P$ symmetry (i.e., $\delta \mu \! = \! 0, h\!\neq\!0$, see \cref{fig:divs_extreme}b), the mapping relations determine a perfectly ``mirrored'' disposition of vertex divergences and pseudo-divergences, with the parabola structures now located in the repulsive sector and the loop structure in the attractive one.

At the same time, as $SU(2)_S$ symmetry is broken for $h\neq 0$, the spin channel splits in the spin transverse (blue lines) channel and in the spin longitudinal channel (green lines), whereas the latter becomes coupled to the charge one (red) in what we have defined as longitudinal sector. In particular, consistent to the mapping, all pairing vertex divergences in the repulsive sector of the $SU(2)_S$ case (yellow lines in \cref{fig:divs_extreme}a) are transformed in the corresponding divergences of the transversal channel in the attractive sector with $SU(2)_P$ symmetry (blue lines in \cref{fig:divs_extreme}b). Indeed, for the same reasons explained above, the generalized susceptibility of the spin transversal channels remains real and bi-symmetric for $\delta\mu \! = \! 0$ when the $SU(2)_P$ symmetry is preserved.
Analogously, all the loop and the parabola structures define now divergences in the combined longitudinal sector, whereas the specific ratio between their components in the charge and spin longitudinal subspaces is highlighted by the red/green color scale\footnote{More quantitatively, one could observe the following: Along a parabola towards the exceptional point the percentage of the longitudinal spin subspace in the divergences increases. 
Remarkably, the maximal mixing happens at the exceptional point. 
Far away from the exceptional point at high $\beta U$ the divergences are almost completely in the longitudinal spin subspace. Note that the pseudo-divergences in the longitudinal are always maximally mixed (i.e. having weight 0.5 in the charge and the spin subspace) if $SU(2)_S$ symmetry is present, which is shown in \cref{eq:max_mix}. }, as detailed in \cref{app:weight}.
Moreover, a parabola in the pairing channel (marked in a thicker yellow line in \cref{fig:divs_extreme}b) is superimposed to every second parabola of divergences in the longitudinal sector.

\subsection{The T=0 limit}

Before coming to the study of the most general case, where both $SU(2)_S$ and $SU(2)_P$ are violated, it is worth to exploit our results to infer how the vertex \mbox{(pseudo-)}divergences evolve in the $T \rightarrow 0$ limit. In fact, in spite of its intrinsic interest (e.g. to discuss the relation between the breakdown of self-consistent perturbation theory and the possible violation of the Luttinger theorem), most of the analysis of the non-perturbative effects on the two-particle level have been restricted to the finite $T$ case.
The choice of working in the dimensionless phase space offers, a valuable opportunity to fill this gap, at least for the HA. Indeed, although our temperature independent representation of the vertex divergences can be used, strictly speaking,  only to analyze finite $T$ cases, the possibility of systematically extracting information for arbitrarily small temperatures and evaluating the asymptotic behavior of vertex \mbox{(pseudo-)}divergence lines allow for a description of the $T=0$ limit.

\begin{figure}[t!]
\includegraphics[scale=0.42]{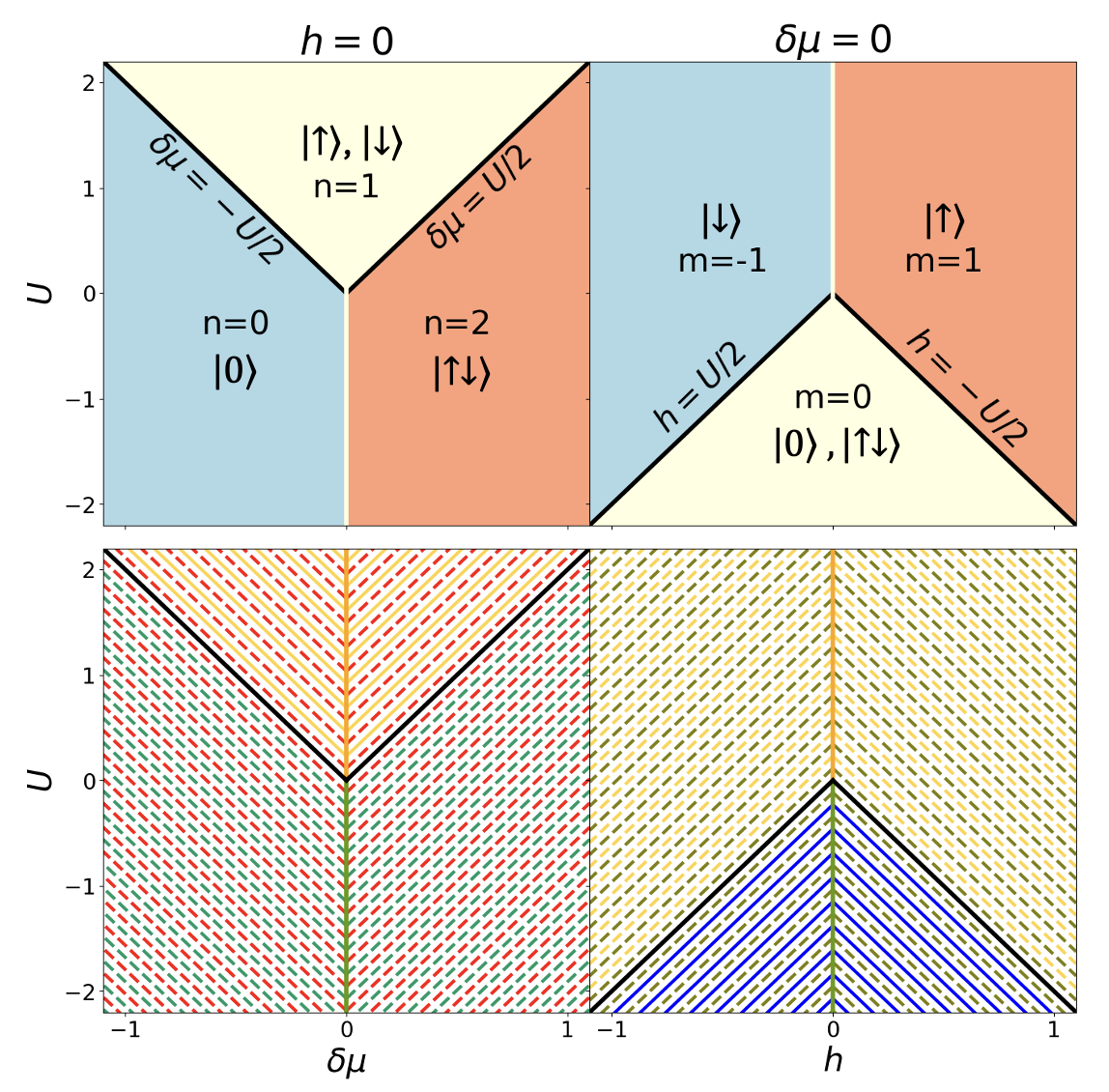}
\caption{\label{fig:divs_extreme_T0} Schematic phase space diagram  of the vertex \mbox{(pseudo-)}divergences \textbf{(bottom panels)} in the Hubbard atom at $T=0$ for the two special cases shown in \cref{fig:divs_extreme} (\textbf{left:} $h=0$ and $\delta\mu\neq 0$; \textbf{right:} $h\neq 0$ and $\delta\mu=0$) compared to the corresponding filling $n=n_\uparrow+n_\downarrow$ \textbf{(upper left)} and magnetization $m=n_\uparrow-n_\downarrow$ \textbf{(upper right)} of the system. These values can be understood by the corresponding ground-state(s) of the HA,  which are explicitly depicted in the different parameter regions.  This schematic phase space diagram should be understood as a continuous distribution of \mbox{(pseudo-)}divergences of different nature in the whole phase space of the HA at $T=0$:  A divergence or pseudo-divergence of the vertex is found at every point in the phase space, whereas the channel in which it occurs is encoded by the color of the dense set of lines shown in the bottom panels. The plotting conventions are as in \cref{fig:divs_extreme}.}
\end{figure}

The results obtained are summarized in the schematic representation of the $T\! = \! = 0$ phase space \cref{fig:divs_extreme_T0}, while more details about the procedure used can be found in \cref{app:T0}.

\begin{figure*}[ht!]
\includegraphics[width=\textwidth]{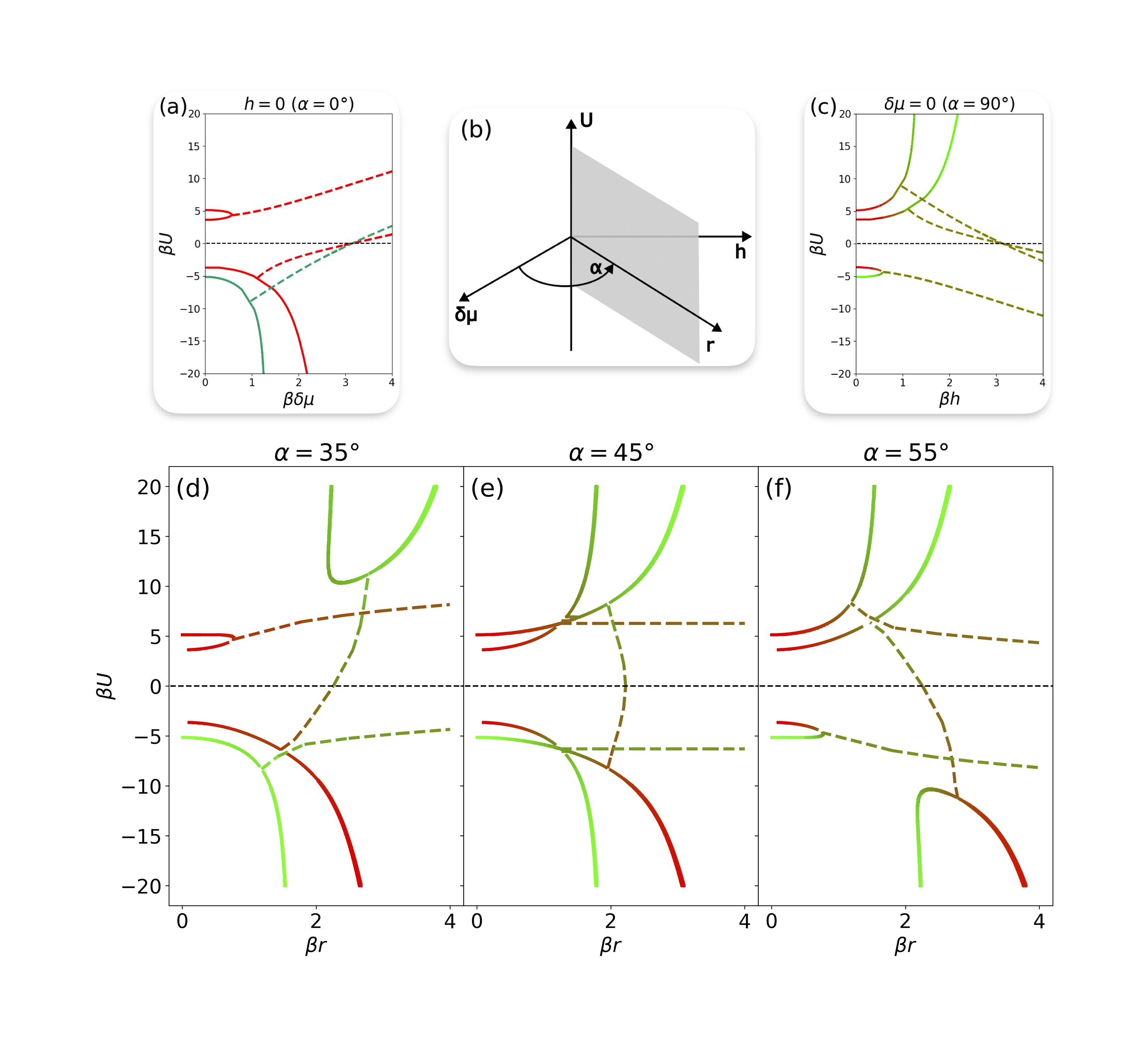}
\vspace{-15mm}
\caption{\label{fig:divs_alpha}
\textbf{(b)}:  Sketch of the new coordinate system $\{U,r,\alpha\}$ introduced to systematically interpolate the illustration of the vertex \mbox{(pseudo-)}divergences of the HA for arbitrary values of $h$ and $\delta \mu$ between the limiting cases shown in \cref{fig:divs_extreme}, which correspond to $\alpha =0^\circ$ and $90^\circ$. The corresponding \mbox{(pseudo-)}divergences in the coupled longitudinal sector already shown in \cref{fig:divs_extreme} are reproduced here, for reference, in \textbf{(a)/(c)}. 
 \textbf{(d)-(f):} Dimensionless phase space diagrams of the vertex \mbox{(pseudo-)}divergences in the coupled longitudinal channel in the most general situation of finite magnetic field and arbitrary filling. Note that only the first divergence line structure is displayed for the different constant $\alpha$ planes chosen (see text). The plotting conventions are as in \cref{fig:divs_extreme}.
}
\end{figure*}

In particular, in the lower panels of \cref{fig:divs_extreme_T0} we observe a {\sl continuous} distribution of vertex divergences at particle-hole symmetry $(\delta\mu=h=0)$, which occur in the charge and pairing(spin) channels for $U>0$ ($U<0$). The continuity of the vertex divergences distribution at $T=0$ evidently reflects the continuity of the Matsubara frequencies in that limit, and appears consistent with the increasingly dense structure of the finite $T$ divergences reported (but not explicitly discussed) in previous works at particle-hole symmetry\cite{schafer2016,thunstrom2018,springer2020}.

\begin{figure*}[ht!]
\includegraphics[width=\textwidth]{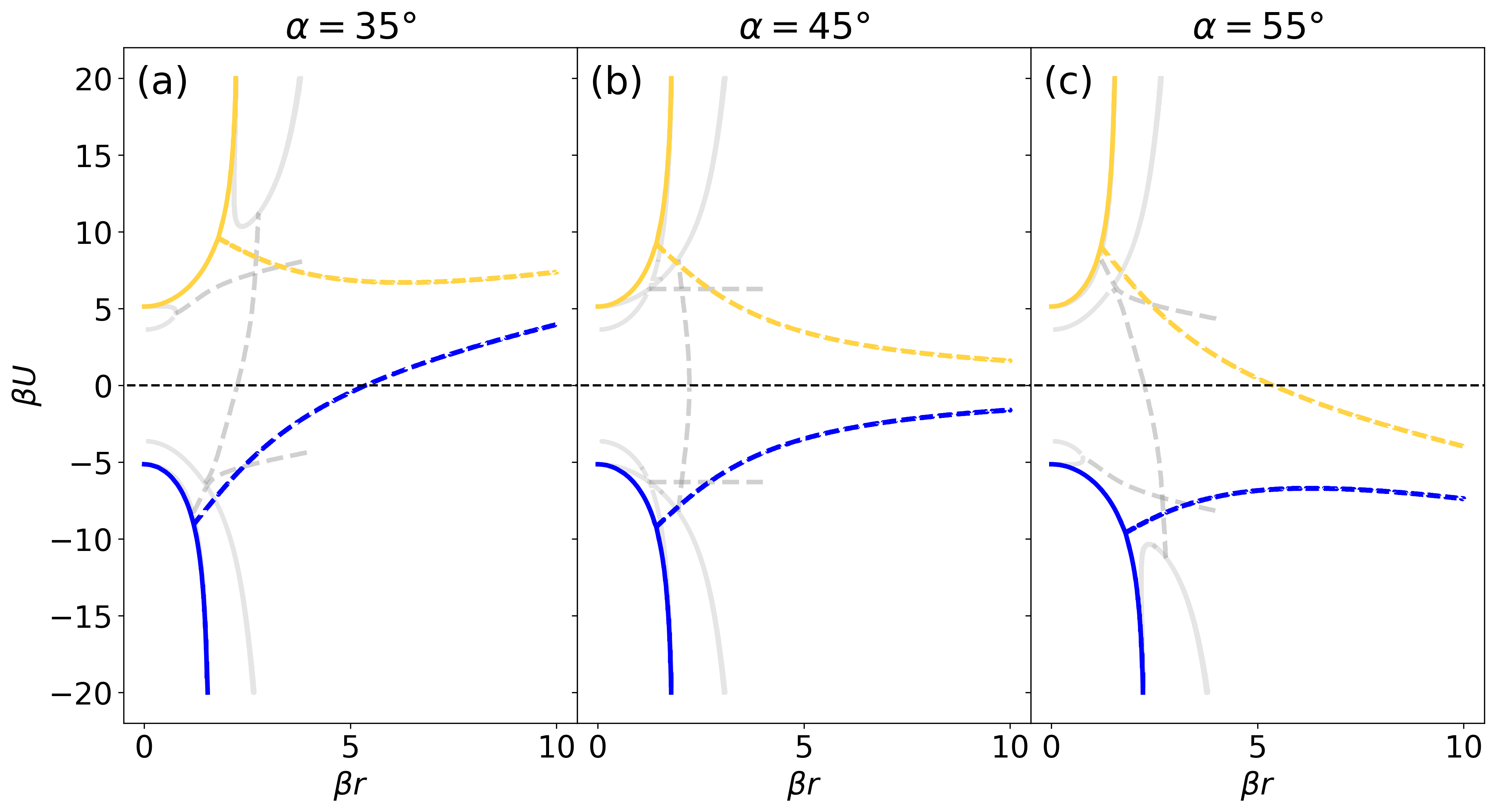}
\caption{\label{fig:divs_transversal} Dimensionless phase space diagrams of the vertex \mbox{(pseudo-)}divergences in the transversal spin channel (blue) and in the pairing channel (yellow) for the Hubbard atom. Note that only the first divergence line structure is displayed for different constant $\alpha$-planes. The vertex \mbox{(pseudo-)}divergences of the coupled longitudinal channel, already shown in \cref{fig:divs_alpha}, are reproduced here in gray for reference.}
\end{figure*}

Of greater interest are the results {\sl out of particle-hole symmetry}. In particular, for $h=0$ (lower left in \cref{fig:divs_extreme_T0}),  we observe  a V-shaped area in the $U>0 $ sector (marked by black lines in \cref{fig:divs_extreme_T0}) located between $\delta\mu=-U/2$ and $\delta\mu=U/2$, which corresponds to the regime of the HA where the ground state is half-filled (see upper left \cref{fig:divs_extreme_T0}). Remarkably, for {\sl every} parameter set within this V-shaped area a vertex divergence in the pairing channel and pseudo-divergence in the charge channel are found simultaneously. Indeed, the latter ones stem from the ``loop-like" structures in \cref{fig:divs_extreme}a, that collapse on the $\delta\mu=0$ axis in the $T=0$ limit. 
Outside the V-shaped area, i.e.~in the region which corresponds to a totally empty or totally full ground-state for all $U$ (see upper left \cref{fig:divs_extreme_T0}),  we find the occurrence of simultaneous pseudo-divergences in the spin and the charge channel everywhere (see \cref{fig:divs_extreme_T0} lower left). The pseudo-divergences originate from the parabola structures in \cref{fig:divs_extreme}a as they progressively move towards the  $\delta \mu=0$-axis in the $U<0$ sector.

In the opposite situation of $\delta \mu=0$ (lower right panel in \cref{fig:divs_extreme_T0}) we find, as expected, a ``mirrored'' distribution of divergences and pseudo-divergences: the V-shaped region is now located in the $U<0$ sector, between $\delta\mu=-U/2$ and $\delta\mu=U/2$. At the same time, due to the violation of the $SU(2)_S$ symmetry and in accordance with the mapping, the V-shaped region entails maximally mixed pseudo-divergences in the longitudinal sector and divergences in the transversal spin channel, while pseudo-divergences in the longitudinal and pairing channel are present outside.

\subsection{Vertex divergences of the HA: The general case}

\begin{figure*}[ht!]
    \includegraphics[width=\textwidth]{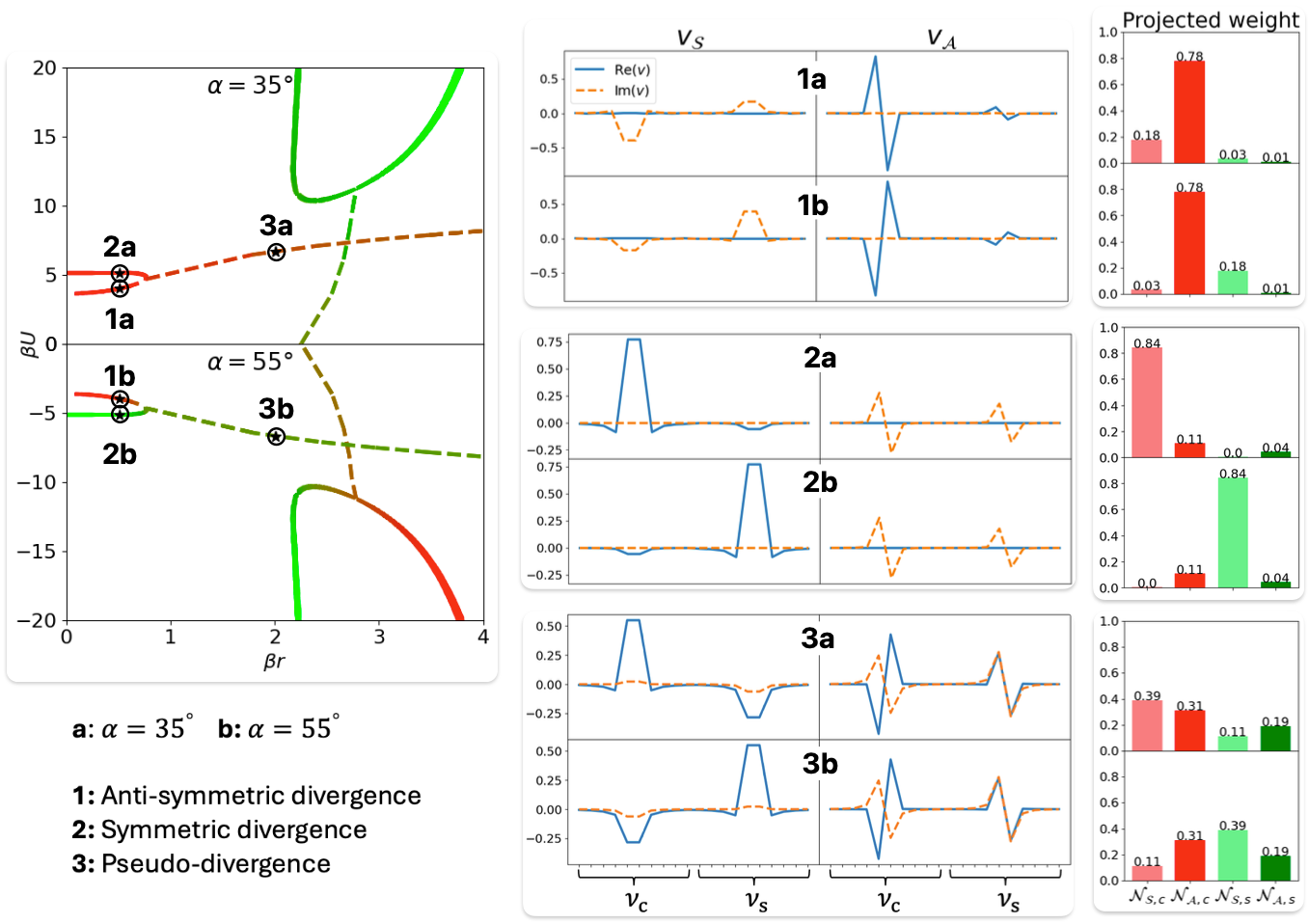}
    \caption{\textbf{Left:} The phase space diagram of the vertex \mbox{(pseudo-)}divergences in longitudinal channel are shown for $\alpha=35^\circ$ in the repulsive sector $U>0$ and for $\alpha=55^\circ$ in the attractive sector $U<0$. These are the two cases that map into each other according the the mapping derived in \cref{sec:der_mapp}. The plotting conventions are as in \cref{fig:divs_extreme}.
    \textbf{Middle:} The eigenvectors that corresponds to the \mbox{(pseudo-)}divergence at specific points, which are marked by black stars. The eigenvectors are split into symmetric $(v_\mathcal{S})$ and antisymmetric part ($v_\mathcal{A}$) to make the mapping more visible. Note that (a) always corresponds to $\alpha=35^\circ$ while (b) corresponds to $\alpha=55^\circ$.
    \textbf{Right:} The projected weights of the eigenvectors again split into symmetric and antisymmetric part.
    }
    \label{fig:divs_mapping}
\end{figure*}

We turn eventually to the discussion of the vertex divergences of the HA in the general case, i.e. for arbitrary values of $h$ and $\delta \mu$.
For the sake of compactness, we introduce  the parameters $\alpha$ and $r$, as sketched in \cref{fig:divs_alpha}b, in order to facilitate the visualization of our results 
 when the values of $\delta\mu, h$ are varied between the limiting cases of $h=0$ ($\alpha =0$) and $\delta \mu=0$ ($\alpha =90^\circ$) shown in \cref{fig:divs_extreme}a and \cref{fig:divs_extreme}b respectively.
 
Specifically, the parameter $\alpha$ and $r$ are defined as the following set of cylinder coordinates:

\begin{align}
\label{eq:cylinder_coordinates}
    \delta \mu = r \cos{\alpha} \quad\text{and}\quad h = r \sin{\alpha}
\end{align}

It is then easy to see that the sets of $\delta \mu, h$ values connected by the mapping linking the attractive and the repulsive model, according to \cref{eq:shibaH}, corresponds to this simple relation for the polar coordinate $\alpha$: 
\begin{align}
    \alpha \leftrightarrow \alpha^\prime \quad\text{with}\quad \alpha + \alpha^\prime = 90^\circ,
\end{align}
whereas $r$ remains unchanged.

Further, due to \cref{eq:flips_chiL,eq:flips_chiSx,eq:flips_chipair}, it is enough to consider $\alpha\in [0^\circ,90^\circ ]$,  as all other regions of the phase space can be reconstructed from this interval.

In \cref{fig:divs_alpha}, we show the evolution of the vertex divergence structures in the longitudinal sector for representative intermediate parameter sets between the limiting situations studied in \cref{sec:limcas}. These correspond to the case of $\alpha =0^\circ$ (\cref{fig:divs_extreme}a) and $\alpha =90^\circ$ (\cref{fig:divs_extreme}b) and  are also reproduced for reference in \cref{fig:divs_alpha}a/c respectively. 
Note that, for the sake of clarity, we only show here the lowest\footnote{In the coupled longitudinal sector for arbitrary $\delta \mu$ and $h$, the labeling of a vertex divergence line with $N$ indicate that the corresponding eigenvector has a maximum on the $N_{th}$ Matsubara frequency in the charge or the spin subspace. Since in \cref{fig:divs_alpha} only the first divergence line structure is displayed, we show in \cref{fig:more_struct} also the second divergence line structure is shown for an exemplary case.} loop and parabola structures and their associated pseudo-divergences.

We consider first the situation where the variation of chemical potential is larger than the magnetic field, i.e. $0^\circ<\alpha<45^\circ$, which we illustrate by showing, as an example, the case of  $\alpha = 35^\circ$ in \cref{fig:divs_alpha}d. Interestingly, for $U>0$, in addition to the loop-structures and their associated pseudo-divergences already discussed for $\alpha =0^\circ$, a parabola structure appears at large values of $\beta U$. We can also observe that the pseudo-divergence associated to this parabola at $U>0$ are connected with the pseudo-divergence emerging from the second parabola for $U<0$.

In the reversed situation, where the magnetic field dominates over $\delta \mu$ (i.e. for $45^\circ<\alpha<90^\circ$), in accordance with the mapping, the phase diagrams are geometrically identical (the projected weights differ as it is predicted by the mapping) as those for $\alpha'= 90^\circ- \alpha$, with $0^\circ<\alpha'<45^\circ$, but for the flipping of the  $U$ and $-U$ sectors (see \cref{fig:divs_alpha}f).

Remarkably, for exactly $\alpha=45^\circ$, where $h$ equals $\delta \mu$, the loop structure touches its corresponding parabola at $U>0$, while the same happens to the pair parabolas for $U<0$.
In both cases, at the point where the two structures touch, a perfectly horizontal (i.e.,~parallel to the $r$-axis) pseudo-divergence line is emerging as it can be seen in \cref{fig:divs_alpha}e.

Our results thus unveil the specific evolution of the vertex (pseudo-)-divergences  between the two qualitatively different situations of $SU(2)_S$ and $SU(2)_P$ symmetries, and, in particular, how the loop-shaped divergence structures gets transformed in the parabola ones and vice-versa.
This also impacts the orientation of the pseudo-divergence lines: One readily notes that those emerging from a parabola always cross the  the $U=0$ axis at $r = \frac{\nu}{\sqrt{1\pm 2 \cos{\alpha}\sin{\alpha}}}$, which is explicitly demonstrated in \cref{eq:pdivU0_uu,eq:pdivU0_dd}, while the ones associated to the loops never cross the non-interacting axis. Evidently, the condition $\alpha=45^\circ$ marks the change between these two behaviors, which is featured by the horizontal pseudo-divergence lines.

Finally, in \cref{fig:divs_transversal} we show for the same three exemplary cases as above (i.e.~$\alpha=35^\circ,45^\circ,55^\circ$) the vertex \mbox{(pseudo-)}divergences in the transversal spin and in the pairing channel, restricting again the plot to the first structures for each channel. The \mbox{(pseudo-)}divergences lines of the longitudinal channel shown in \cref{fig:divs_mapping} are replotted in gray for reference.

All the divergence lines in these channels display a parabolic shape, consistent to the fact that,  at particle-hole symmetry, only symmetric divergences for both transversal channels  exist. However, different from the longitudinal sector, the pseudo-divergences associated to the transversal spin channel cross the $U=0$ axis at $r = \frac{\nu}{\sqrt{(\cos{\alpha})^2-(\sin{\alpha})^2}}$, i.e.~only for $0^\circ\leq \alpha < 45^\circ$, while those of the pairing channel cross the $U=0$ axis at $r = \frac{\nu}{\sqrt{(\sin{\alpha})^2-(\cos{\alpha})^2}}$ and therefore only for $45^\circ < \alpha \leq 90^\circ$. 

\subsection{The mapping of the eigenvectors}

Crucial information~\cite{gunnarsson2017,springer2020,vanloon2020,reitner2020,reitner2023,kowalski2023} on the possible physical effects associated to divergences is encoded in the frequency structure of the associated eigenvectors, and, especially, in their overall symmetry properties. While a detailed discussion of these properties is reported in \cref{app:sus_syms}, which generalizes the corresponding particle-hole symmetric results of \cite{springer2020}, here we illustrate some essential aspects of the eigenvector behavior, by hand of a pertinent example.

To this aim, we consider the following two cases, directly related by the mapping, i.e., $\alpha=35^\circ$ for $U>0$ and $\alpha=55^\circ$ for $U<0$ (see left panel of \cref{fig:divs_mapping}).
The eigenvectors corresponding to the (pseudo-)divergences of three selected parameter sets (marked by black stars in the left panel) are then shown in the central panel of the figure, where each eigenvector is split into its symmetric ($v_\mathcal{S}$) and and antisymmetric $(v_\mathcal{A})$ part, while their projected weight in the charge and spin subspace (see \cref{app:weight}), calculated for the symmetric and antisymmetric part of each eigenvector, is shown in the rightmost panel.

According to the relations in \cref{sec:der_mapp},  the eigenvector marked by \textbf{1a} is mapped into \textbf{1b}, \textbf{2a} is mapped into \textbf{2b} and \textbf{3a} into \textbf{3b}.
When considering \textbf{1a} and \textbf{1b}, which are associated to an antisymmetric divergence of the vertex, we can observe the characteristic symmetry properties of the eigenvectors corresponding to a real eigenvalue in a $\kappa$-real matrix $\chi_\text{L}$ (see \cref{eq:kappa_real_ev}). In particular, these eigenvectors split into purely real anti-symmetric part and imaginary symmetric part. Analogously, \textbf{2a} and \textbf{2b}, associated to a symmetric divergence display purely real symmetric/purely imaginary anti-symmetric parts.
On the other hand, \textbf{3a} and \textbf{3b} is one of the eigenvectors associated to a pseudo-divergence, and hence, to a pair of complex conjugate eigenvalues. Therefore, the symmetric and anti-symmetric parts of these eigenvectors are no longer purely real or imaginary.
Note that only one of the eigenvectors of the pseudo-divergence is shown since the other one can be reconstructed by \cref{eq:kappa_real_ev} and the projected weights of these two eigenvectors are the same.

Eventually, by comparing the \textbf{a} and \textbf{b} figures of each case, one can easily recognize the behavior predicted by the mapping relations derived in the previous section. The symmetric part of the eigenvector $v_\mathcal{S}$ changes from charge to spin subspace and vice versa, while the anti-symmetric part $v_\mathcal{A}$ remains invariant under the mapping. This also reflects into the relations of the corresponding weights, as it is readily seen when projecting the portion of the symmetric/antisymmetric part of the eigenvectors onto the charge or the spin subspace.

\subsection{Finite bosonic transfer frequency}
\label{sec:finite_w}

According to the state-of-the-art knowledge (cf.~Ref.~\cite{gunnarsson2017}) only irreducible vertex divergences occurring at $\omega=0$ can be directly linked to the breakdown of the self-consistent perturbation theory for the many-electron problem. For this reason, the illustrative study of the (pseudo-)divergences of the HA presented in this work is mostly focused on the $\omega=0$ case.

Notwithstanding, for the sake of completeness, it is worth to get a glimpse on some relevant modifications of the results presented so far, when considering  vertex divergences structures at finite transfer frequency ($\omega \neq 0$).
To this aim, we will focus our discussion of the finite $\omega$-case on the (pseudo-)divergences occurring in the charge and spin sectors, when $SU(2)_S$ symmetry is preserved (i.e., $\delta\mu\neq 0, h=0$). In this respect, our starting point will be represented by the findings of Ref.~\cite{thunstrom2018}, where vertex divergences of the particle-hole symmetric HA ($\delta \mu = h = 0$) have been shortly analyzed for  specific finite frequency values.

More precisely, the analytical  expressions of Ref.~\cite{thunstrom2018} show that, for the ph-symmetric HA, vertex divergences occur at all $\beta U$ values such that (i)  $U = \pm \frac{2}{\sqrt{3}}\sqrt{\nu (\nu+\omega)}$ in the charge channel or (ii) \mbox{$U = \pm 2\sqrt{-\nu (\nu+\omega)}$} in the spin channel, where $\nu=\frac{(2n-1)\pi}{\beta}$ and $\omega=\frac{2m \pi}{\beta}$ are the fermionic and the bosonic Matsubara frequencies, respectively. The eigenvectors associated to these divergences display an antisymmetric structure in the \emph{shifted} frequency space \cref{eq:freq_shift}, i.e. they are proportional to $\delta_{\nu,\nu^\prime}-\delta_{\nu,-\nu^\prime-\omega}$.

Since the discriminant in the above relations (i) and (ii) must be positive, anti-symmetric divergences in the spin channel can only occur for $\omega \neq 0$, when $\nu(\nu+\omega)<0$. At the same time, for these frequency values  no divergence can occur in the charge channel.
However, the fulfillment of  $\nu(\nu+\omega)<0$ is not a sufficient condition for an anti-symmetric divergence to occur, since anti-symmetric eigenvectors require two distinct fermionic frequencies $\nu\neq -\nu-\omega$ for their construction. As a result, divergences in the spin channel only appear if $|m|>1$ and a new anti-symmetric divergence only appears at even $m$ values.

Furthermore, also vertex divergences associated to symmetric eigenvectors in the shifted frequency space can be found \cite{thunstrom2018}, namely for all $U$ values where
\begin{align}
\label{eq:sym_div_omega}
    f_r = \frac{U\tan{(\frac{\beta}{4}(\sqrt{4B^2_r+\omega^2}+\omega))}}{\sqrt{4B^2_r+\omega^2}}\pm 1= 0
\end{align}
is fulfilled, where the $+$ ($-$) sign is for the charge (spin) channel and

\begin{align}
\label{eq:B_r}
    B_\text{c} = \frac{U}{2}\sqrt{\frac{3e^{\beta U/2}-1}{1+e^{\beta U/2}}},\\
    B_\text{s} = -\frac{U}{2}\sqrt{\frac{3e^{-\beta U/2}-1}{1+e^{-\beta U/2}}}.
\end{align}

\begin{figure}[t!]
\includegraphics[scale=0.6]{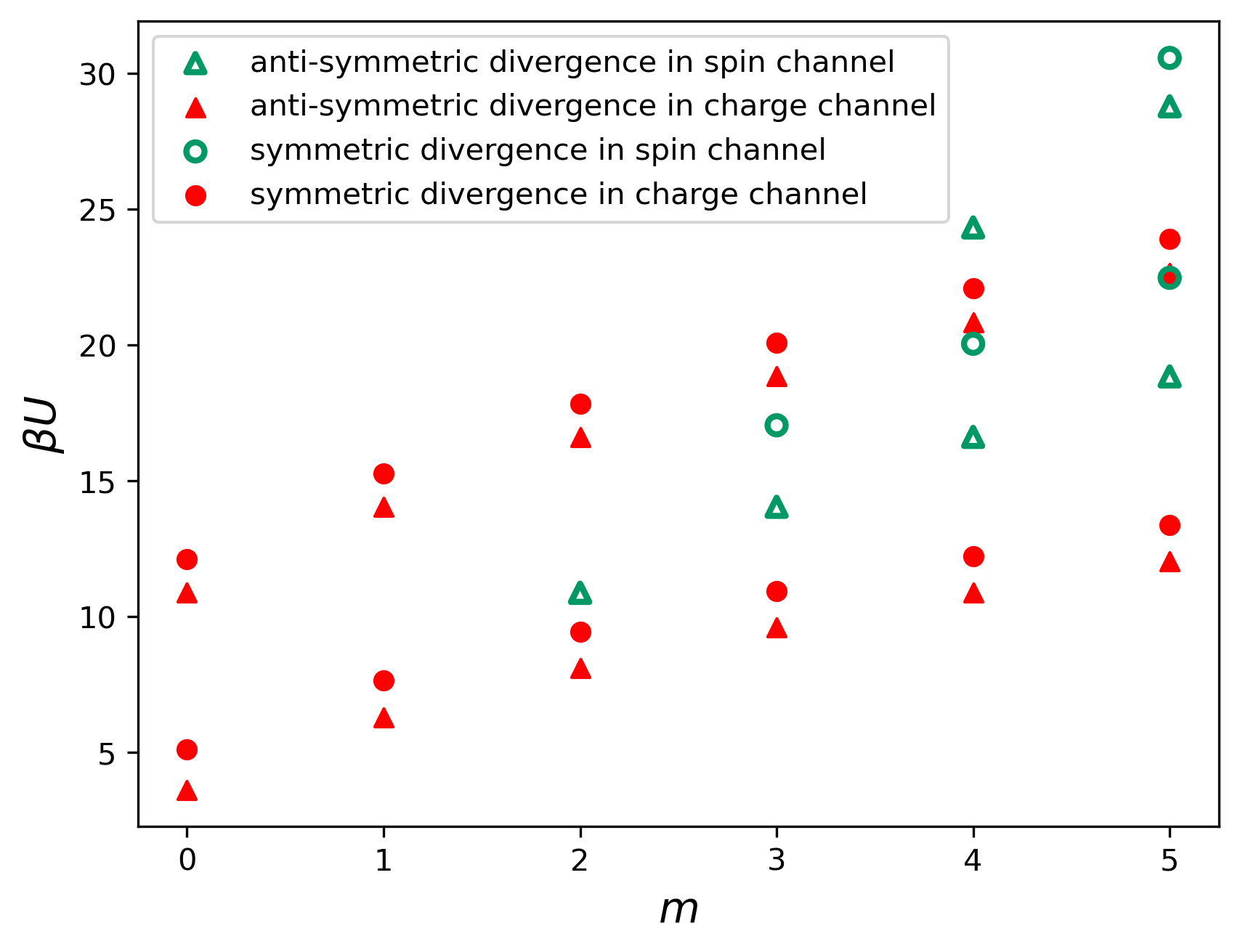}
\caption{\label{fig:finite_omega} Location of the first two symmetric and two anti-symmetric vertex divergences in charge and spin channel at $ph$-symmetry for different $\omega = 2\pi m /\beta$ values.}
\end{figure}

From \cref{eq:sym_div_omega} one finds~\cite{thunstrom2018} the occurrence of symmetric divergences in the charge channel for all values of $\omega$ if $U>0$. However, we should note that the condition of \cref{eq:sym_div_omega} can not be fulfilled, if the corresponding eigenvector is maximal at a $\nu$ value, for which $\nu(\nu+\omega)<0$. Note that this is the same condition as for the anti-symmetric divergences.

As for the symmetric divergences in the spin channel, differently as reported in a side remark in Ref.~\cite{thunstrom2018}, these can indeed occur at $U>0$ for $m\geq 3$.

In summary, in the spin channel we find $\lfloor m/2\rfloor$ anti-symmetric divergences, while an explicit evaluation of \cref{eq:sym_div_omega} also yields $\lceil m/2\rceil-1$ symmetric divergences. 
At the same time, in the charge channel, infinitely many symmetric and anti-symmetric divergences occur, under the condition that the corresponding eigenvectors do not have a maximum at frequencies where $\nu(\nu+\omega)<0$.

The location of the first two symmetric and two anti-symmetric divergences in charge and spin channel is shown in \cref{fig:finite_omega} for different values of $\omega$ at ph-symmetry. The location of the divergences is symmetric in $\omega$, i.e. it is the same for $\omega$ and $-\omega$.

For particle-hole symmetry the action of the Shiba mapping at finite $\omega$ can be evidenced by noting that (i) all antisymmetric vertex divergences occur at the same absolute $U$ values for positive and negative $U$, and (ii) that $f_\text{c}$ is transformed into $-f_\text{s}$ by changing $U$ to $-U$. The latter condition implies that the symmetric vertex divergences for $U<0$ can be obtained by just exchanging  charge and spin channel in the discussion for $U>0$.
Hence,  as a direct consequence of the Shiba mapping at finite frequency see \cref{eq:chi_shiba_main}, the findings of Ref.~\cite{springer2020} for the vertex divergence of the static particle-hole case remain identically applicable also at $\omega \neq 0$.

At the same time, a remarkable difference to the $\omega=0$ case is that negative real eigenvalues of the generalized charge/spin susceptibility at $ph$-symmetry can be found already at $U=0$ (see \cref{eq:chi0_ph}) namely when the frequency condition $\nu(\nu+\omega)<0$ is fulfilled.
However, the role played by these negative eigenvalues in triggering additional divergences at $\omega \neq 0$ is quite different in the charge and in the spin channel. 
In fact, in the former one, consistently with the discussion above, the negative eigenvalues found at $U=0$ never cross zero by increasing $U$.
In the spin channel, instead, \emph{all} negative eigenvalues found at $U=0$  will cross zero by increasing $U$, except the ones that correspond to eigenvectors which either have their maximum at the central frequency $\nu^*=-\omega/2$ if $m$ is odd or that are symmetric and have their maximum at $\nu=\omega/2\pm\pi/\beta$ if $m$ is even. Such crossings evidently trigger vertex divergences in the spin sector that were \emph{not} present for $\omega=0$.

Eventually, merging the description of the vertex divergences at finite $\omega$ for the ph-symmetric case~\cite{thunstrom2018} with the knowledge of the $\delta \mu$-evolution of the vertex divergences at $\omega=0$ (\cref{sec:limcas}), it becomes easy to intuitively outline their qualitative behavior as a function of a varying chemical potential.
Specifically, by gradually increasing the absolute value of
$\delta\mu$, we find that pairs of symmetric and anti-symmetric vertex divergences get closer in parameter space,  forming \emph{loop}-structures in the phase diagram analogous to those shown in Fig.~\ref{fig:divs_extreme}. In contrast, vertex divergences that do not have a partner at $ph$-symmetry, i.e. there exists no vertex divergence which has its maximum at the same frequency but has opposite symmetry, form a \emph{parabola}-like structure in the phase space. Such divergences are, for example, the anti-symmetric divergences in the spin channel for $m=2,4$ at the largest $\beta U$ value shown in \cref{fig:finite_omega}. 

Hence, the overall evolution of the vertex (pseudo-)divergence lines for $\omega\neq 0$ follows the intuitive expectation gained from the $\omega=0$ case. These intuitive picture is quantitatively confirmed by direct calculations (not shown).

\section{Conclusions}
\label{sec:conclusio}

In this work we have derived the explicit relations mapping the on-site two-particle generalized susceptibilities of bipartite lattice models with repulsive local interaction into those of corresponding models with attractive interaction and vice versa. The presented derivations, which exploit the Shiba transformation on the two-particle level (i) at arbitrary filling and/or in the presence of a finite magnetic field, as well as (ii) for finite transfer (bosonic) Matsubara frequency, extend the results of previous studies~\cite{springer2020}, which were restricted to the case of static correlation functions in the specific, and less realistic situation of perfect particle-hole symmetry of the system.

Considering the increasing importance played by the calculation and the manipulation of the two-particle quantities~\cite{rohringer2012,hafermann2014,kugler2021,lee2021} for different cutting-edge algorithms~\cite{thunstrom2018,tagliavini2018,lee2021} designed for treating strong correlations in many-electron systems, the explicit derivation of the mapping relations on the two-particle level might be useful in several contexts. For instance, it may allow to reduce the computational effort of numerically heavy two-particle calculations in challenging parameter regimes and provide rigorous benchmark testbeds for computational schemes based on generalized on-site susceptibilities and vertex functions beyond the special case of particle hole-symmetry. Further, on a more fundamental level, our findings might represent a useful guide for an improved understanding of the information encoded in the correlation functions on the two-particle level. 

In the latter respect, as a pertinent example for the applicability of the mapping relations, we have systematically analyzed location and nature of the divergences of the irreducible vertex functions in the Hubbard atom with repulsive and attractive on-site interaction for arbitrary filling and finite magnetic field, overcoming the parameter restrictions of preceding studies \cite{schafer2016,thunstrom2018}.
The evident symmetries characterizing our dimensionless phase space diagrams for the divergences in the HA represent a direct consequence of the derived mapping relations. This consideration allows, in turn, for an easier comprehension of the intrinsic links existing between the various divergence structures in different parameter regimes of the HA. 

Further, among the properties characterizing the evolution of the singularities in the different scattering channels as a function of filling and magnetic field, it may be  worth to stress here the appearance of the exceptional points (EPs) at which the vertex divergences (associated to a vanishing eigenvalue of the corresponding generalized susceptibility) transform into pseudo-divergences (associated to a complex conjugate eigenvalue pair with zero real part).
In fact, the relevance of these EPs has been recently discussed in the specific context of a possible topological protection \cite{reitner2023} of the thermodynamic phase-separation instabilities emerging \cite{kotliar2002,eckstein2007,vanloon2020,reitner2020,kowalski2023} in the proximity of a Mott-Hubbard metal-to-insulator transitions of correlated lattice systems. At the same time, as demonstrated in Ref.~\cite{vucicevic2018}, their occurrence is {\sl not} limited to purely on-site four-point correlation functions. Eventually, the EPs might play, together with the pseudo-divergences discussed in our work as well as with the pole structures \cite{rohshap2024} possibly appearing in the complex interaction plane\cite{wu2017}, an important role in controlling the convergence properties of the self-consistent perturbation expansions for the many-electron problem \cite{kozik2015,stan2015,gunnarsson2017,thunstrom2018}, {\sl beyond} \cite{vucicevic2018}  the rather specific condition of perfect particle-hole symmetry condition mostly investigated so far.

\begin{acknowledgments}
We want to thank Sabine Andergassen, Massimo Capone, Sergio Ciuchi, Lorenzo Crippa, Lorenzo Del Re, Dominik Fus, Marcel Gievers, Anna Kauch, Erik van Loon, Georg Rohringer, Stefan Rohshap and Thomas Sch\"afer for insightful discussions. We acknowledge financial support from the Austrian Science Fund (FWF) via the projects I 5487 (A.T.) and I 5868, project P01 of the research unit FOR 5249  of the German research foundation (DFG), (H.E.); M. R. acknowledges support as a recipient of a DOC fellowship of the Austrian Academy of Sciences. G.S. acknowledges support from the Deutsche Forschungsgemeinschaft (DFG, German Research Foundation) through QUAST FOR 5249 (Project No. 449872909, project P5). Calculations have been partly performed on the Vienna Scientific Cluster.
\end{acknowledgments}

\appendix
\section{Symmetries of generalized susceptibilities}
\label{app:sus_syms}

In this section, we study the properties of the local generalized susceptibilities for the symmetries considered in the main text, with a particular focus on $SU(2)_S$, $SU(2)_P$, and $SO(4)$ symmetry. For a detailed derivation of these symmetry properties and an in-depth discussion, the reader is referred to Ref.~\cite{rohringer2013}.

Doing a full particle-hole transformation ($c_\sigma \to c^\dagger_\sigma$, $c^\dagger_\sigma \to c_\sigma$) on the Hamiltonian in \cref{eq:shibaH} gives

\begin{align}
    H(U,\delta\mu,h) \xleftrightarrow{\text{ph-trafo}} H(U,-\delta\mu,-h).
\end{align}

Therefore, the system is particle-hole symmetric (i.e., the Hamiltonian is invariant under the particle-hole transformation) if $\delta\mu\!=\!h\!=\!0$, which corresponds to $SO(4)$ symmetry.

Further, applying a full particle-hole transformation in the definitions of \cref{eq:gensus,eq:gensus_bar} and then using crossing symmetry and complex conjugation (\cref{eq:chi_sym_cc,eq:chi_sym_cs}) lead to

\begin{align}
\label{eq:chi_sig_ph_trafo}
    \chi \xleftrightarrow{\text{ph-trafo}}\chi^*,
\end{align}
and a sign change in the magnetic field just flips all spins of the generalized susceptibility $(\sigma\rightarrow-\sigma, \; \forall \sigma)$.

We find the following relations for the different channels:
In the coupled longitudinal channel

\begin{align}
\label{eq:flips_chiL}
\begin{split}
    \chi_\text{L} \xleftrightarrow{\delta\mu, h \leftrightarrow -\delta\mu, -h} \chi_\text{L}^*, \\
    \chi_\text{L} \xleftrightarrow{h \leftrightarrow -h} M\chi_\text{L}M, \\
    \chi_\text{L} \xleftrightarrow{\delta\mu \leftrightarrow -\delta\mu} M\chi_\text{L}^*M,
\end{split}
\end{align}
where $M = \begin{pmatrix}
        \mathbb{1} & \mathbb{0}\\
        \mathbb{0} & -\mathbb{1}
    \end{pmatrix}$, and for the transversal spin and the pairing channel we find

\begin{align}
\label{eq:flips_chiSx}
\begin{split}
    \chi_{S_x} \xleftrightarrow{\delta\mu, h \leftrightarrow -\delta\mu, -h} \chi_{S_x}^*, \\
    \chi_{S_x}\xleftrightarrow{h \leftrightarrow -h} \chi_{S_x}, \\
    \chi_{S_x} \xleftrightarrow{\delta\mu \leftrightarrow -\delta\mu} \chi_{S_x}^*,
\end{split}
\end{align}

\begin{align}
\label{eq:flips_chipair}
\begin{split}
    \chi_\text{pair} \xleftrightarrow{\delta\mu, h \leftrightarrow -\delta\mu, -h} \chi_\text{pair}^*, \\
    \chi_\text{pair} \xleftrightarrow{h \leftrightarrow -h} \chi_\text{pair}^*, \\
    \chi_\text{pair} \xleftrightarrow{\delta\mu \leftrightarrow -\delta\mu} \chi_\text{pair}.
\end{split}
\end{align}

Since the transformations in \cref{eq:flips_chiL,eq:flips_chiSx,eq:flips_chipair} change neither the eigenvalues nor the projected weights (see \mbox{\cref{eq:phys_weight})} of the generalized susceptibility matrices, we confirm that the whole phase space can be generated by mapping the calculated quantities from the sub-region where $\delta\mu>0$ and $h>0$. This corresponds to the interval $\alpha \in [0^\circ
,90^\circ]$ in the cylinder coordinates of \cref{eq:cylinder_coordinates}.

In \cref{eq:chi_sym_cs,eq:chi_sym_cc,eq:chi_sym_HinR,eq:chi_sym_su2,eq:chi_sym_su2p} the properties of the generalized susceptibilities under the specific symmetry relations are explicitly written following Ref.~\cite{rohringer2013}.

Crossing symmetry and complex conjugation are fundamental ``symmetries", i.e. they are always present, $H\in\mathbb{R}$ refers to the Hamiltonian being a real function of $c$ and $c^\dagger$ and $SU(2)_S$/$SU(2)_P$ means symmetric  with respect to spin/pseudo-spin.

\begin{align}
\label{eq:chi_sym_cs}
\begin{split}
    \text{Crossing Symmetry: } \chi^{\nu,\nu^\prime\! ,\,\omega}_{\sigma\sigma^\prime,\text{ph}}&=\chi^{\omega+\nu^\prime\! ,\,\omega+\nu,\,-\omega}_{\sigma^\prime\sigma,\text{ph}},\\
    \chi^{\nu,\nu^\prime\! ,\,\omega}_{\overline{\sigma\sigma^\prime},\text{ph}}&=\chi^{\omega+\nu^\prime\! ,\,\omega+\nu,\,-\omega}_{\overline{\sigma^\prime\sigma},\text{ph}},\\
    \chi^{\nu,\nu^\prime\! ,\,\omega}_{\overline{\sigma\sigma^\prime},\text{pp}}&=\chi^{\omega-\nu\! ,\,\omega-\nu^\prime\! ,\,\omega}_{\overline{\sigma^\prime\sigma},\text{pp}},
\end{split}
\end{align}

\begin{align}
\label{eq:chi_sym_cc}
\begin{split}
    \text{Complex Conjugation: } \chi^{\nu,\nu^\prime\! ,\,\omega}_{\sigma\sigma^\prime,\text{ph}}&=\Big(\chi^{-\nu^\prime\! ,\,-\nu-,\,-\omega}_{\sigma^\prime\sigma,\text{ph}}\Big)^*,\\
    \chi^{\nu,\nu^\prime\! ,\,\omega}_{\overline{\sigma\sigma^\prime},\text{ph}}&=\Big(\chi^{-\nu^\prime\! ,\,-\nu-,\,-\omega}_{\overline{\sigma\sigma^\prime},\text{ph}}\Big)^*,\\
    \chi^{\nu,\nu^\prime\! ,\,\omega}_{\overline{\sigma\sigma^\prime},\text{pp}}&=\Big(\chi^{-\nu^\prime\! ,\,-\nu-,\,-\omega}_{\overline{\sigma\sigma^\prime},\text{pp}}\Big)^*,
\end{split}
\end{align}

\begin{align}
\label{eq:chi_sym_HinR}
\begin{split}
    \hat{H} \in \mathbb{R} \text{: } \chi^{\nu,\nu^\prime\! ,\,\omega}_{\sigma\sigma^\prime,\text{ph}}&=\chi^{\nu^\prime\! ,\,\nu,\,\omega}_{\sigma^\prime\sigma,\text{ph}},\\
    \chi^{\nu,\nu^\prime\! ,\,\omega}_{\overline{\sigma\sigma^\prime},\text{ph}}&=\chi^{\nu^\prime\! ,\,\nu,\,\omega}_{\overline{\sigma\sigma^\prime},\text{ph}},\\
    \chi^{\nu,\nu^\prime\! ,\,\omega}_{\overline{\sigma\sigma^\prime},\text{pp}}&=\chi^{\nu^\prime\! ,\,\nu,\,\omega}_{\overline{\sigma\sigma^\prime},\text{pp}},
\end{split}
\end{align}

\begin{align}
\label{eq:chi_sym_su2}
\begin{split}
    \text{$SU(2)_S$ symmetry: } \chi^{\nu,\nu^\prime\!,\,\omega}_{\sigma\sigma^\prime,\text{ph}}&=\chi^{\nu,\nu^\prime\!,\,\omega}_{-\sigma-\sigma^\prime,\text{ph}},\\
    \chi^{\nu,\nu^\prime\!,\,\omega}_{\overline{\sigma\sigma^\prime},\text{ph}}&=\chi^{\nu,\nu^\prime\!,\,\omega}_{\overline{-\sigma-\sigma^\prime},\text{ph}},\\
    \chi^{\nu,\nu^\prime\!,\,\omega}_{\overline{\sigma\sigma^\prime},\text{pp}}&=\chi^{\nu,\nu^\prime\!,\,\omega}_{\overline{-\sigma-\sigma^\prime},\text{pp}},
\end{split}
\end{align}

\begin{align}
\begin{split}
\label{eq:chi_sym_su2p}
    \text{$SU(2)_P$ symmetry: } \chi^{\nu,\nu^\prime\!,\,\omega}_{\sigma\sigma^\prime,\text{ph}}&=\Big(\chi^{\nu,\nu^\prime\!,\,\omega}_{-\sigma-\sigma^\prime,\text{ph}}\Big)^*,\\
    \chi^{\nu,\nu^\prime\!,\,\omega}_{\overline{\sigma\sigma^\prime},\text{ph}}&=\Big(\chi^{\nu,\nu^\prime\!,\,\omega}_{\overline{-\sigma-\sigma^\prime},\text{ph}}\Big)^*,\\
    \chi^{\nu,\nu^\prime\!,\,\omega}_{\overline{\sigma\sigma^\prime},\text{pp}}&=\Big(\chi^{\nu,\nu^\prime\!,\,\omega}_{\overline{-\sigma-\sigma^\prime},\text{pp}}\Big)^*.
\end{split}
\end{align}

From \cref{eq:chi_sym_cs,eq:chi_sym_cc,eq:chi_sym_HinR,eq:chi_sym_su2,eq:chi_sym_su2p} the relations written in \cref{tab:symtable} of the main text can be derived. 

In the following, we introduce the class of centro-Hermitian matrices $C$ which fulfill

\begin{align}
\label{eq:centro_hermitian}
     J C^* J = C,\quad\text{with }J=\mqty(\admat{1,\udots,1}).
\end{align}

Centro-Hermitian matrices have eigenvalues that are either real or come in complex conjugate pairs \cite{lee1980}.

By using the symmetry \cref{eq:chi_sym_cs,eq:chi_sym_cc,eq:chi_sym_HinR} we find that all generalized susceptibilities are centro-Hermitian for $\omega=0$. 
Note that for $\chi_{\sigma\sigma^\prime,\text{ph}}^{\omega=0}$ the centro-Hermitian property is fundamental, since only complex conjugation and crossing symmetry (which are always present) are needed, whereas for $\chi_{\overline{\sigma\sigma^\prime}\text{ph/pp}}^{\omega=0}$ also  $H\in\mathbb{R}$ is required (which could be violated in principle). 
Without this restriction $\chi_{\overline{\sigma\sigma^\prime}\text{ph/pp}}^{\omega=0}$ are per-hermitian matrices \cite{reitner2023}.

Moreover, $\chi_{\sigma\sigma^\prime,\text{ph}}$ can be considered as a centro-Hermitian matrix for finite $\omega$ when shifting the fermionic Matsubara frequencies by $\nu^{(\prime)}\rightarrow \nu^{(\prime)}+\lfloor m/2\rfloor 2\pi/\beta$ with $\omega = 2\pi m/\beta$ ($\lfloor x \rfloor$ refers to $\operatorname{floor}(x)$). 
This leads to 

\begin{align}
\label{eq:freq_shift}
    \chi^{\nu,\nu^\prime\! ,\,\omega}_{\sigma\sigma^\prime,\text{ph}}&=\Big(\chi^{-\nu-\omega\! ,\,-\nu^\prime-\omega,\,\omega}_{\sigma\sigma^\prime,\text{ph}}\Big)^* \nonumber\\
    &\Bigg\downarrow \text{frequency shift}\\
    \chi^{\nu-\lfloor\omega/2\rfloor,\nu^\prime-\lfloor\omega/2\rfloor ,\,\omega}_{\sigma\sigma^\prime,\text{ph}}&=\Big(\chi^{-\nu-\lfloor\omega/2\rfloor ,\,-\nu^\prime-\lfloor\omega/2\rfloor,\,\omega}_{\sigma\sigma^\prime,\text{ph}}\Big)^*, \nonumber
\end{align}
where $\lfloor\omega/2\rfloor$ is a short-hand notation for $\lfloor m/2\rfloor 2\pi/\beta$ such that $\nu-\omega/2$ is still a fermionic Matsubara frequency. 
Thus, for any bosonic frequency $\omega$ $\chi_{\sigma\sigma^\prime,\text{ph}}$ is a centro-Hermitian matrix in the frequency shifted space.

Moreover, we introduce the class of $\kappa$-real matrices \cite{hill1992}, which also have only real or complex conjugate pairs as eigenvalues. The generalized susceptibility of the coupled longitudinal channel $\chi_\mathrm{L}$ is a $\kappa$-real matrix which fulfills the relation

\begin{align}
\label{eq:kappa_real}
     \Pi K^* \Pi = K \quad\text{with}\quad \Pi = \begin{pmatrix}
         J & \mathbb{0}\\
         \mathbb{0} & J
     \end{pmatrix}.
\end{align}

Further, all considered channels are symmetric matrices if $H\in\mathbb{R}$ and can be diagonalized by a complex orthogonal transformation, if there is no exceptional point.
The corresponding inner product of the orthogonal Euclidean quasi-norm is defined as

\begin{align}
\label{eq:T_scalar_prod}
    \langle u, v\rangle := u^T\cdot v = \sum_i^n u_i v_i
\end{align}
and
\begin{align}
\label{eq:T_norm}
    \norm{u}^2_T:=u^T \cdot u.
\end{align}

Eigenvectors of symmetric and centro-Hermitian matrices that are normalized with respect to \cref{eq:T_norm} have the following properties \cite{reitner2023}:

\begin{align}
\label{eq:centro_hermitian_ev}
\begin{split}
     &\text{if } \lambda_\alpha\in\mathbb{R} \text{ (and not degenerate):}\; v_\alpha = \pm J v_\alpha^*,\\
     &\text{if } \lambda_\alpha\in\mathbb{C}:\; \exists v_{\alpha^\prime} = J v_\alpha^*,\\
\end{split}
\end{align}
where the relation for real eigenvalues leads to the fact that their eigenvectors have a symmetric/antisymmetric real part with an antisymmetric/symmetric imaginary part (with respect to the shifted Matsubara frequency space $\nu^{(\prime)}\rightarrow \nu^{(\prime)}+\lfloor m/2\rfloor 2\pi/\beta$).

For the symmetric $\kappa$-real matrix $\chi_\text{L}$ we can make  a similar argument as in Ref.~\cite{reitner2023} to obtain:

\begin{align}
\label{eq:kappa_real_ev}
\begin{split}
     &\text{if } \lambda_\alpha\in\mathbb{R} \text{ (and not degenerate):}\; v_\alpha = \pm \Pi v_\alpha^*,\\
     &\text{if } \lambda_\alpha\in\mathbb{C}:\; \exists v_{\alpha^\prime} = \Pi v_\alpha^*,\\
\end{split}
\end{align}
where the relation for real eigenvalues leads to the fact that their eigenvectors have a symmetric/antisymmetric real part with an antisymmetric/symmetric imaginary part in the charge and spin subspaces respectively (with respect to the shifted Matsubara frequency space).

\section{BSE of the transversal channel}
\label{app:BSE_transversal}

In general, $\chi_{S_x}$ and $\chi_\text{pair}$ do not represent decoupled channels in the BSE. Even though $\chi_{S_x}=\chi_{S_y}$, they still couple to each other for $\langle S_z\rangle \neq 0$ \cite{delre2021}. The same considerations apply to the real and imaginary part of the pairing field $\Delta$, which are given by $S_{p,x}$ and $S_{p,y}$, and couple for $\langle S_{p,z} \rangle \neq 0$. Therefore the $S_x$/pairing channel generally (for arbitrary $\omega$) only decouples within the transversal subspace when $SU(2)_{S/P}$ symmetry is fulfilled. This can be shown from the symmetry relations in \cref{app:sus_syms} and the expressions of the coupled channels in \cref{eq:chiT,eq:chiTpp}. This is an example of how the mapping ``preserves" the decoupling of the channels regarding the BSE.
Instead the transverse components $\chi_{\overline{\sigma\sigma^\prime}}$ and $\chi_{\overline{\sigma\sigma^\prime},\text{pp}}$ remain independent in the BSE. Hence, for the respective  generalized susceptibility of the coupled BSE we have to consider the transversal ph-channel (or $\chi_{\overline{\sigma\sigma^\prime}}$ directly)

\begin{align}
\label{eq:chiT}
    \chi_\text{T} = 
    \begin{pmatrix}
            \chi_{\overline{\uparrow\downarrow}}+\chi_{\overline{\downarrow\uparrow}} & \chi_{\overline{\uparrow\downarrow}}-\chi_{\overline{\downarrow\uparrow}}\\
            \chi_{\overline{\uparrow\downarrow}}-\chi_{\overline{\downarrow\uparrow}} & \chi_{\overline{\uparrow\downarrow}}+\chi_{\overline{\downarrow\uparrow}}
    \end{pmatrix}
\end{align}
instead of the spin channel and
the transversal pp-channel (or $\chi_{\overline{\sigma\sigma^\prime},\text{pp}}$ directly)

\begin{align}
\label{eq:chiTpp}
    \chi_\text{T,pp} = 
    \begin{pmatrix}
            -\chi_{\overline{\uparrow\downarrow},\text{pp}}-\chi_{\overline{\downarrow\uparrow},\text{pp}}^* & -\chi_{\overline{\uparrow\downarrow},\text{pp}}+\chi_{\overline{\downarrow\uparrow},\text{pp}}^*\\
            -\chi_{\overline{\uparrow\downarrow}}+\chi_{\overline{\downarrow\uparrow},\text{pp}}^* & -\chi_{\overline{\uparrow\downarrow},\text{pp}}-\chi_{\overline{\downarrow\uparrow},\text{pp}}^*
    \end{pmatrix}
\end{align}
instead of the pairing channel.
If the Hamiltonian is a real function of $c$ and $c^\dagger$ then $\chi_{\overline{\uparrow\downarrow}}^{\nu,\nu^\prime\! ,\,\omega}=\chi_{\overline{\downarrow\uparrow}}^{\omega+\nu,\omega+\nu^\prime\! ,\,-\omega}$ and $\chi_{\overline{\uparrow\downarrow},\text{pp}}^{\nu,\nu^\prime\! ,\,\omega}=(\chi_{\overline{\downarrow\uparrow},\text{pp}}^{\nu-\omega,\nu^\prime-\omega ,\,\omega})^*$ (see  \cref{app:sus_syms}), and both the BSE of
the pairing channel and the transversal spin channel decouple for $\omega=0$. Hence, to investigate the divergences of the two-particle
irreducible vertices for $\omega=0$, $\chi_{S_x}$ and $\chi_\text{pair}$  can be used for the scope of this paper as it is done in \cref{sec:application}.

\section{Calculations for the Shiba mapping of the generalized susceptibilities}
\label{app:mapping}

In this section, we derive the action of the Shiba transformation applied to the local generalized susceptibilities. The results of this transformation are shown in \cref{sec:der_mapp}. 
For the derivation we start by showing how the orthogonal transformation 

\begin{equation}
     Q = \frac{1}{\sqrt{2}}    
    \begin{pmatrix}
    \mathbb{1} & -J\\
    \mathbb{1} & J
    \end{pmatrix}
\end{equation}
separates the symmetric and antisymmetric parts of a centro-Hermitian matrix. 
Every centro-Hermitian matrix $\chi_\text{CH}$ can be represented by the sum of a centro-symmetric real matrix $\chi^\prime$ ($J\chi^\prime J=\chi^\prime$) and a skew-centro-symmetric imaginary matrix $\chi^{\prime\prime}$ ($J\chi^{\prime\prime} J=-\chi^{\prime\prime}$).

For even dimensions a centro-Hermitian matrix can be represented by

\begin{align}
\begin{split}
    \chi_\text{CH}=\chi^\prime+i\chi^{\prime\prime}\quad\text{with}\quad \chi^\prime &= 
    \begin{pmatrix}
        A^\prime & JB^\prime J\\
        B^\prime & JA^\prime J
    \end{pmatrix},\\
    \chi^{\prime\prime} &= 
    \begin{pmatrix}
        A^{\prime\prime} & -JB^{\prime\prime} J\\
        B^{\prime\prime} & -JA^{\prime\prime} J
    \end{pmatrix},
\end{split}
\end{align}
where $\chi^{\prime(\prime)}$, $A^{\prime(\prime)}$ and $B^{\prime(\prime)}$ are real square matrices. 

The transformation $Q$ acts on (skew-)centro-symmetric matrices as follows

\begin{align}
    Q\chi^\prime Q^T = 
    \begin{pmatrix}
        A^\prime-JB^\prime & \mathbb{0}\\
        \mathbb{0} &  A^\prime+JB^\prime 
    \end{pmatrix}
\end{align}
and
\begin{align}
Q\chi^{\prime\prime}Q^T = 
    \begin{pmatrix}
        \mathbb{0} & A^{\prime\prime}-JB^{\prime\prime}\\
        A^{\prime\prime}+JB^{\prime\prime} & \mathbb{0}
    \end{pmatrix}.
\end{align}
$A^\prime-JB^\prime$ and $A^{\prime\prime}+JB^{\prime\prime}$ are referred to as antisymmetric blocks and $A^\prime+JB^\prime$ and $A^{\prime\prime}-JB^{\prime\prime}$ are termed symmetric blocks. To understand why, we let $\chi^\prime$ and $\chi^{\prime\prime}$ act on a generic vector $v$ which is decomposed into symmetric and antisymmetric parts:

\begin{align}
\begin{split}
    \chi^{\prime} v
    &=\begin{pmatrix}
        A^\prime & JB^\prime J\\
        B^\prime & JA^\prime J
    \end{pmatrix}\left[
    \begin{pmatrix}
        v_\mathcal{S}\\
        Jv_\mathcal{S}
    \end{pmatrix}+
    \begin{pmatrix}
        v_\mathcal{A}\\
        -Jv_\mathcal{A}
    \end{pmatrix}\right]\\
    &=
    \begin{pmatrix}
        (A^\prime+JB^\prime)v_\mathcal{S} +(A^\prime-JB^\prime)v_\mathcal{A}\\
        (C^\prime+JB^\prime)v_\mathcal{S} +(C^\prime-JB^\prime)v_\mathcal{A}
    \end{pmatrix}
\end{split}
\end{align}
and
\begin{align}
\begin{split}
    \chi^{\prime\prime} v
    &=\begin{pmatrix}
        A^{\prime\prime} & -JB^{\prime\prime} J\\
        B^{\prime\prime} & -JA^{\prime\prime} J
    \end{pmatrix}\left[
    \begin{pmatrix}
        v_\mathcal{S}\\
        Jv_\mathcal{S}
    \end{pmatrix}+
    \begin{pmatrix}
        v_\mathcal{A}\\
        -Jv_\mathcal{A}
    \end{pmatrix}\right]\\
    &=
    \begin{pmatrix}
        (A^{\prime\prime}-JB^{\prime\prime})v_\mathcal{S} +(A^{\prime\prime}+JB^{\prime\prime})v_\mathcal{A}\\
        (B^{\prime\prime}-JA^{\prime\prime})v_\mathcal{S} +(B^{\prime\prime}+JA^{\prime\prime})v_\mathcal{A}
    \end{pmatrix}.
\end{split}
\end{align}

As it can be seen $A^\prime-JB^\prime$ and $A^{\prime\prime}+JB^{\prime\prime}$ couple to the antisymmetric part of the vector and $A^\prime+JB^\prime$ and $A^{\prime\prime}-JB^{\prime\prime}$ couple to the symmetric part of the vector.

Therefore we identify the blocks in \cref{eq:q_trafo_ch} by

\begin{align}
\begin{split}
    \chi_\mathcal{A}^\prime = A^\prime-JB^\prime,\\
    \chi_\mathcal{S}^{\prime} = A^\prime+JB^\prime,\\
    \chi_\mathcal{A}^{\prime\prime} = A^{\prime\prime}+JB^{\prime\prime},\\
    \chi_\mathcal{S}^{\prime\prime} = A^{\prime\prime}-JB^{\prime\prime}.
\end{split}
\end{align}

As described in \cref{app:sus_syms} the generalized susceptibility $\chi_{\sigma\sigma^\prime,\text{ph}}$ is still a centro-Hermitian matrix for $\omega\neq 0$ if we consider the shifted space $\nu^{(\prime)}\rightarrow \nu^{(\prime)}+\lfloor m/2\rfloor 2\pi/\beta$ with $\omega = 2m\pi/\beta$. 
As already mentioned in the main text the derivation for even $m$ is completely analogous to the $\omega=0$ case if the shifted space is considered. 

However, for odd $m$ slight adaptations are needed since the matrix acquires an odd frequency structure with a new central frequency $\nu^* = -\omega/2$. In other words, if we consider a finite frequency box (e.g., in numerical calculations) the frequency shift leads to a $(2N)\times (2N)$ matrix for even $m$ and to a $(2N+1)\times (2N+1)$ matrix for odd $m$.

A centro-Hermitian matrix of odd dimension can be depicted by
\begin{align}
\begin{split}   \chi_\text{CH}=\chi^\prime+i\chi^{\prime\prime},\quad\text{with}\quad \chi^\prime &= 
    \begin{pmatrix}
        A^\prime & a^\prime & JB^\prime J\\
        b^\prime & c & b^\prime J\\
        B^\prime & Ja^\prime & JA^\prime J
    \end{pmatrix},\\
    \quad \chi^{\prime\prime} &= 
    \begin{pmatrix}
        A^{\prime\prime} & a^{\prime\prime} & -JB^{\prime\prime} J\\
        b^{\prime\prime} & 0 & -b^{\prime\prime} J\\
        B^{\prime\prime} & -Ja^{\prime\prime} & -JA^{\prime\prime} J
    \end{pmatrix},
\end{split}
\end{align}
where $A,B \in \mathbb{R}^{N\times N}$, $a \in \mathbb{R}^{N\times 1}$, $b \in \mathbb{R}^{1\times N}$ and $c \in \mathbb{R}$. 
With
\begin{align}
\label{eq:q_trafo_shift}
    Q = \frac{1}{\sqrt{2}}    
    \begin{pmatrix}
    \mathbb{1} & 0 & -J\\
    0&\sqrt{2}&0\\
    \mathbb{1} &0 & J
    \end{pmatrix}
\end{align}
we can apply the transformation $Q\chi_\text{CH}Q^T$, in the same way as for the even-dimensional case to identify the symmetric and anti-symmetric blocks of the matrix in \cref{eq:q_trafo_ch}:

\begin{align}
\begin{split}
    \chi_\mathcal{A}^\prime = A^\prime-JB^\prime,\\
    \chi_\mathcal{S}^{\prime} = 
    \begin{pmatrix}
        c&\sqrt{2}b^\prime\\
        \sqrt{2}a^\prime&A^\prime+JB^\prime\\
    \end{pmatrix},\\
    \chi_\mathcal{A}^{\prime\prime} = 
    \begin{pmatrix}
        \sqrt{2}b^{\prime\prime}\\
        A^{\prime\prime}+JB^{\prime\prime}
    \end{pmatrix},\\
    \chi_\mathcal{S}^{\prime\prime} = 
    \begin{pmatrix}
        \sqrt{2}a^{\prime\prime},&A^{\prime\prime}-JB^{\prime\prime} 
    \end{pmatrix}.
\end{split}
\end{align}

Now we can start with the derivation of the mapping of the generalized susceptibilities by writing the action of the Shiba transformation in \cref{eq:chi_shiba} as matrix products in the frequency shifted space:

\begin{align}
\label{eq:chi_shiba_matrix}
\begin{split}
    \chi_{\uparrow\uparrow,\text{U}} = \chi_{\uparrow\uparrow,\text{-U}}, \; \;\chi_{\downarrow\downarrow,\text{U}} = J\chi_{\downarrow\downarrow,\text{-U}}J, \\ \chi_{\uparrow\downarrow,\text{U}} = -\chi_{\uparrow\downarrow,\text{-U}}J, \; \;\text{and}\;\;\chi_{\downarrow\uparrow,\text{U}} = -J\chi_{\downarrow\uparrow,\text{-U}}.
\end{split}
\end{align}

With these building blocks, we can construct the coupled longitudinal matrix

\begin{equation}
\chi_\text{L} =     
\begin{pmatrix}
\chi_\text{c} & \chi_\text{cs}\\
\chi_\text{sc} & \chi_\text{s}
\end{pmatrix}
\end{equation}
 and then apply the transformation 
\begin{equation}
    \mathcal{Q}= 
    \begin{pmatrix}
        Q & \mathbb{0}\\
        \mathbb{0} & Q
    \end{pmatrix} 
\end{equation}
to it.

With simple matrix multiplication, we obtain

\begin{align}
\label{eq:chi_full_shiba}
\begin{split}
    \mathcal{Q}\chi_\text{L,-U}&\mathcal{Q}^T= 
    \begin{pmatrix}
        Q\chi_\text{c,-U}Q^T & Q\chi_\text{cs,-U}Q^T\\
        Q\chi_\text{sc,-U}Q^T & Q\chi_\text{s,-U}Q^T
    \end{pmatrix}\\
    &=\begin{pmatrix}
        \chi_{\text{c,-U},\mathcal{A}}^\prime & i\chi_{\text{c,-U},\mathcal{S}}^{\prime\prime}& \chi_{\text{cs,-U},\mathcal{A}}^\prime & i\chi_{\text{cs,-U},\mathcal{S}}^{\prime\prime}\\
        i\chi_{\text{c,-U},\mathcal{A}}^{\prime\prime} & \chi_{\text{c,-U},\mathcal{S}}^{\prime}&i\chi_{\text{cs,-U},\mathcal{A}}^{\prime\prime} & \chi_{\text{cs,-U},\mathcal{S}}^{\prime}\\
        \chi_{\text{sc,-U},\mathcal{A}}^\prime & i\chi_{\text{sc,-U},\mathcal{S}}^{\prime\prime}&\chi_{\text{s,-U},\mathcal{A}}^\prime & i\chi_{\text{s,-U},\mathcal{S}}^{\prime\prime}\\
        i\chi_{\text{sc,-U},\mathcal{A}}^{\prime\prime} &\chi_{\text{sc,-U},\mathcal{S}}^{\prime}&i\chi_{\text{s,-U},\mathcal{A}}^{\prime\prime} &\chi_{\text{s,-U},\mathcal{S}}^{\prime}
    \end{pmatrix}\\
    &=\begin{pmatrix}
        \chi_{\text{c,U},\mathcal{A}}^\prime & i\chi_{\text{cs,U},\mathcal{S}}^{\prime\prime}& \chi_{\text{cs,U},\mathcal{A}}^\prime & i\chi_{\text{c,U},\mathcal{S}}^{\prime\prime}\\
        i\chi_{\text{sc,U},\mathcal{A}}^{\prime\prime} & \chi_{\text{s,U},\mathcal{S}}^{\prime}&i\chi_{\text{s,U},\mathcal{A}}^{\prime\prime} & \chi_{\text{sc,U},\mathcal{S}}^{\prime}\\
        \chi_{\text{sc,U},\mathcal{A}}^\prime & i\chi_{\text{s,U},\mathcal{S}}^{\prime\prime}&\chi_{\text{s,U},\mathcal{A}}^\prime & i\chi_{\text{sc,U},\mathcal{S}}^{\prime\prime}\\
        i\chi_{\text{c,U},\mathcal{A}}^{\prime\prime} &\chi_{\text{cs,U},\mathcal{S}}^{\prime}&i\chi_{\text{cs,U},\mathcal{A}}^{\prime\prime} &\chi_{\text{c,U},\mathcal{S}}^{\prime}
    \end{pmatrix}\\
    &=T\mathcal{Q}\chi_\text{L,U}\mathcal{Q}^TT,
\end{split}
\end{align}
where an appropriate $Q$ matrix has to be used depending on whether the generalized susceptibility is an even or an odd matrix. From \cref{eq:chi_full_shiba} and the fact that $\mathcal{Q}$ is an orthogonal transformation \cref{eq:chi_shiba_main} in the main text is obtained.

In \cref{eq:chi_full_shiba} one can see that after the $\mathcal{Q}$ transformation $\chi_\text{L,U}$ and $\chi_\text{L,-U}$ only differ by a permutation of the different sub matrices. This reshuffling is done by the matrix $T$ in \cref{eq:QandT_trafo} which needs to be adapted to \cref{eq:T_trafo_shift} for the odd-dimensional case

\begin{align}
\label{eq:T_trafo_shift}
    T=
    \begin{pmatrix}
        \mathbb{1}&0&\mathbb{0}&\mathbb{0}&0&\mathbb{0}\\
        0&0&0&0&1&0&\\
        \mathbb{0}&0&\mathbb{0}&\mathbb{0}&0&\mathbb{1}\\
        \mathbb{0}&0&\mathbb{0}&\mathbb{1}&0&\mathbb{0}\\
        0&1&0&0&0&0&\\
        \mathbb{0}&0&\mathbb{1}&\mathbb{0}&0&\mathbb{0}\\
    \end{pmatrix}.
\end{align}

We find that (i), a symmetric/anti-symmetric block of $\chi_\text{L,U}$ always
maps to a symmetric/anti-symmetric block of $\chi_\text{L,-U}$ and (ii), all $\chi_\mathcal{A}^\prime$ blocks are invariant under the mapping. All other blocks get redistributed according to \cref{eq:chi_full_shiba}, which is summarized in \cref{eq:mapping_rules_re,eq:mapping_rules_im} in the main text.\footnote{The $SO(4)$ symmetric case of Ref.~\cite{springer2020} can be obtained by noticing that for $SU(2)_P$ symmetry the matrices have no imaginary part and for $SU(2)_S$ symmetry $\chi_\text{cs}$ and $\chi_\text{sc}$ vanish.} 
Note that for $\chi_\mathrm{L}$ the subscript $\mathcal{A/S}$ refers to the coupling of the respective anti-/symmetric components in the charge and spin subspace of a generic vector $v$.
For the even-dimensional case

\begin{align}
    v =  \begin{pmatrix}
        v_{\text{c},\mathcal{S}}\\
        Jv_{\text{c},\mathcal{S}}\\
        v_{\text{s},\mathcal{S}}\\
        Jv_{\text{s},\mathcal{S}}
    \end{pmatrix} +
    \begin{pmatrix}
        v_{\text{c},\mathcal{A}}\\
        -Jv_{\text{c},\mathcal{A}}\\
        v_{\text{s},\mathcal{A}}\\
        -Jv_{\text{s},\mathcal{A}}
    \end{pmatrix}
\end{align}
and for the odd-dimensional case

\begin{align}
    v =  \begin{pmatrix}
        v_{\text{c},\mathcal{S}}\\
        m_\text{c}\\
        Jv_{\text{c},\mathcal{S}}\\
        v_{\text{s},\mathcal{S}}\\
        m_\text{s}\\
        Jv_{\text{s},\mathcal{S}}
    \end{pmatrix} +
    \begin{pmatrix}
        v_{\text{c},\mathcal{A}}\\
        0\\
        -Jv_{\text{c},\mathcal{A}}\\
        v_{\text{s},\mathcal{A}}\\
        0\\
        -Jv_{\text{s},\mathcal{A}}
    \end{pmatrix}.
\end{align}

Eventually, we consider the action of the Shiba transformation $S = \mathcal{Q}^T T\mathcal{Q}$, which is orthogonal and symmetric, on the eigenvectors and eigenvalues of $\chi_\mathrm{L}$.

Since $S$ is a orthogonal transformation, $\chi_\text{L,U}$ and $\chi_\text{L,-U}$ have the same eigenvalues. Further, if $v$ is a eigenvector of $\chi_\text{L,U}$ then $Sv$ is an eigenvector of $\chi_\text{L,-U}$ with the same eigenvalue. Transforming the eigenvector $v$ we find

\begin{align}
\begin{split}
\label{eq:map_ev_even}
    Sv=S&\left[
    \begin{pmatrix}
        v_{\text{c},\mathcal{S}}\\
        Jv_{\text{c},\mathcal{S}}\\
        v_{\text{s},\mathcal{S}}\\
        Jv_{\text{s},\mathcal{S}}
    \end{pmatrix} +
    \begin{pmatrix}
        v_{\text{c},\mathcal{A}}\\
        -Jv_{\text{c},\mathcal{A}}\\
        v_{\text{s},\mathcal{A}}\\
        -Jv_{\text{s},\mathcal{A}}
    \end{pmatrix}
    \right] \\
    &= 
    \begin{pmatrix}
        v_{\text{s},\mathcal{S}}\\
        Jv_{\text{s},\mathcal{S}}\\
        v_{\text{c},\mathcal{S}}\\
        Jv_{\text{c},\mathcal{S}}
    \end{pmatrix} +
    \begin{pmatrix}
        v_{\text{c},\mathcal{A}}\\
        -Jv_{\text{c},\mathcal{A}}\\
        v_{\text{s},\mathcal{A}}\\
        -Jv_{\text{s},\mathcal{A}}
    \end{pmatrix}
\end{split}
\end{align}
for the even dimensional case and

\begin{align}
\begin{split}
\label{eq:map_ev_odd}
    Sv=S&\left[
    \begin{pmatrix}
        v_{\text{c},\mathcal{S}}\\
        m_\text{c}\\
        Jv_{\text{c},\mathcal{S}}\\
        v_{\text{s},\mathcal{S}}\\
        m_\text{s}\\
        Jv_{\text{s},\mathcal{S}}
    \end{pmatrix} +
    \begin{pmatrix}
        v_{\text{c},\mathcal{A}}\\
        0\\
        -Jv_{\text{c},\mathcal{A}}\\
        v_{\text{s},\mathcal{A}}\\
        0\\
        -Jv_{\text{s},\mathcal{A}}
    \end{pmatrix}
    \right] \\
    &= 
    \begin{pmatrix}
        v_{\text{s},\mathcal{S}}\\
        m_\text{s}\\
        Jv_{\text{s},\mathcal{S}}\\
        v_{\text{c},\mathcal{S}}\\
        m_\text{c}\\
        Jv_{\text{c},\mathcal{S}}
    \end{pmatrix} +
    \begin{pmatrix}
        v_{\text{c},\mathcal{A}}\\
        0\\
        -Jv_{\text{c},\mathcal{A}}\\
        v_{\text{s},\mathcal{A}}\\
        0\\
        -Jv_{\text{s},\mathcal{A}}
    \end{pmatrix}
\end{split}
\end{align}
for the odd-dimensional case. Evidently, the symmetric part of the eigenvector changes the subspace between charge and spin, while the anti-symmetric part of the eigenvector is invariant under the transformation.

To obtain the Shiba transformation for the generalized susceptibility in the transverse spin channel $\chi_{S_x}$ we apply the Shiba transformation on \cref{eq:gensus_bar} and get

\begin{align}
\begin{split}
    \chi_{\overline{\uparrow\downarrow},\text{U}}^{\nu,\nu^\prime\! ,\,\omega} &= -\chi_{\overline{\uparrow\downarrow},\text{-U}}^{\nu,\nu^\prime\! ,\,-\nu-\nu^\prime-\omega} = -\chi_{\overline{\uparrow\downarrow},\text{pp,-U}}^{\nu,\nu^\prime\! ,\,-\omega} ,\\
    \chi_{\overline{\downarrow\uparrow},\text{U}}^{\nu,\nu^\prime\! ,\,\omega} &= -\chi_{\overline{\downarrow\uparrow},\text{-U}}^{-\nu,-\nu^\prime\! ,\,\nu+\nu^\prime+\omega} = -\chi_{\overline{\downarrow\uparrow},\text{pp,-U}}^{-\nu,-\nu^\prime\! ,\,\omega} = -\Big(\chi_{\overline{\downarrow\uparrow},\text{pp,-U}}^{\nu,\nu^\prime\! ,\,-\omega} \Big)^*,
\end{split}
\end{align}
where we use the fact that shifting $\omega\rightarrow \omega-\nu-\nu^\prime$ is changing the ph-convention to the pp-convention and, in the second line, we applied the property for complex conjugation \cref{eq:chi_sym_cc}.

Together with the definition of $\chi_{S_x}^{\nu,\nu^\prime\! ,\,\omega}$ and $\chi_\text{pair}^{\nu,\nu^\prime\! ,\,\omega}$ in \cref{eq:channels_sx,eq:channels_end} we get

\begin{align}
    \chi_{S_x,\text{U}}^{\nu,\nu^\prime\! ,\,\omega}=\chi_\text{pair,-U}^{\nu,\nu^\prime\! ,\,-\omega},
\end{align}
which is, in our notation, equivalent to \cref{eq:map_bar} in the main text.

\section{Analytical expression for the generalized susceptibility in the Hubbard atom}
\label{app:sus_HA}

Following Ref.~\cite{pairault2000}, the one-particle Green's function and the connected two-particle Green's functions for the HA in ph-notation for different spin combinations are as follows:

\begin{align}
\label{eq:g1}
    G_{1,\sigma}(\nu)=\frac{1-n_{-\sigma}}{i\nu+\mu+\sigma h}+\frac{n_{-\sigma}}{i\nu+\mu+\sigma h -U},
\end{align}

\onecolumngrid
\begin{align}
\label{eq:g2upup}
    G_{2,\uparrow\uparrow}^\text{con}(\nu, \nu^\prime, \omega)=\frac{\beta U^2 n_\downarrow(1-n_\downarrow)(\delta_{\omega0}-\delta_{\nu\nu^\prime})}{(i\nu+i\omega+\mu+ h)(i\nu+i\omega+\mu+ h-U)(i\nu^\prime+\mu+ h)(i\nu^\prime+\mu+ h-U)},
\end{align}

\begin{align}
\label{eq:g2_updown}
\begin{aligned}
&G_{2,\uparrow \downarrow}^\text{con}\left(\nu, \nu^{\prime}, \omega\right)\\
&=\frac{n_\uparrow+n_\downarrow-1}{i\nu+ i\nu^{\prime}+i\omega+2 \mu-U}\left(\frac{1}{i\nu+i\omega+\mu+h-U}+\frac{1}{i \nu^{\prime}+\mu-h-U}\right)\left(\frac{1}{i\nu^{\prime}+i\omega+\mu-h-U}+\frac{1}{i \nu+\mu+h-U}\right)\\
&+\frac{n_\uparrow-n_\downarrow}{i\nu^\prime-i\nu-2h}\left(\frac{1}{i\nu^\prime+\mu-h}-\frac{1}{i\nu+\mu+h-U}\right)\left(\frac{1}{i \nu^\prime + i\omega+\mu-h}-\frac{1}{i \nu+i\omega+\mu+h-U}\right)\\
&+\frac{\beta U^{2} \delta_{\omega 0}\left(e^{(2 \mu-U) \beta}-e^{2 \mu \beta}\right)}{\left(1+e^{(\mu+h) \beta} + e^{(\mu-h) \beta} +e^{(2 \mu-U) \beta}\right)^{2}} \frac{1}{(i\nu+\mu+h)(i\nu+\mu+h-U)\left(i \nu^{\prime}+\mu-h\right)\left(i \nu^{\prime}+\mu-h-U\right)}\\
&+\frac{n_\uparrow-1}{(i\nu^\prime+i\omega+\mu-h)\left(i\nu^{\prime}+\mu-h\right)(i \nu+\mu+h-U)}+\frac{1-n_\uparrow}{(i\nu+i\omega+\mu+h-U)\left(i \nu+\mu+h-U\right)\left(i\nu^{\prime}+\mu-h\right)}\\
&+\frac{1-n_\downarrow}{(i\nu+i\omega+\mu+h)\left(i \nu^{\prime}+\mu-h-U\right)\left(i\nu+\mu+h-U\right)}+\frac{n_\downarrow-1}{\left(i \nu+i\omega+\mu+h\right)\left(i\nu+\mu+h-U\right)(i \nu^\prime+i\omega+\mu-h)}\\
&+\frac{1-n_\downarrow}{(i\nu^\prime+\mu-h)\left(i \nu+i\omega+\mu+h\right)(i \nu^\prime+i\omega+\mu-h)}+\frac{1-n_\downarrow}{(i\nu^\prime+\mu-h)\left(i \nu+i\omega+\mu+h\right)\left(i\nu+\mu+h\right)}\\
&+\frac{1-n_\downarrow}{(i\nu^\prime+\mu-h-U)\left(i\nu+\mu+h\right)(i \nu^\prime+i\omega+\mu-h-U)}+\frac{n_\downarrow-1}{(i\nu^\prime+\mu-h-U)\left(i \nu+i\omega+\mu+h\right)\left(i\nu+\mu+h\right)}\\
&+\frac{1-n_\downarrow}{\left(i \nu+i\omega+\mu+h-U\right)\left(i\nu+\mu+h\right)(i \nu^\prime+i\omega+\mu-h-U)}+\frac{n_\downarrow-1}{(i\nu^\prime+\mu-h)\left(i \nu+i\omega+\mu+h-U\right)\left(i\nu+\mu+h\right)}\\
&+\frac{-n_\uparrow}{(i\nu^\prime+\mu-h-U)\left(i \nu+i\omega+\mu+h-U\right)(i \nu^\prime+i\omega+\mu-h-U)}+\frac{-n_\uparrow}{(i\nu^\prime+\mu-h-U)\left(i \nu+i\omega+\mu+h-U\right)\left(i\nu+\mu+h-U\right)},
\end{aligned}
\end{align}
\twocolumngrid
where 

\begin{align}
    n_\sigma = \frac{e^{(\mu+\sigma h)\beta}+e^{(2\mu-U)\beta}}{1+e^{(\mu+h) \beta} + e^{(\mu-h) \beta} +e^{(2 \mu-U) \beta}}
\end{align}
and $G_{2,\downarrow\downarrow}^\text{con}$ and $G_{2,\downarrow\uparrow}^\text{con}$ can be obtained from \cref{eq:g2upup} and \cref{eq:g2_updown} by changing $h$ to $-h$.
Note that the terms in the first two lines in \cref{eq:g2_updown} can become singular (i.e. proportional to $\delta_{\nu,\nu^\prime+\omega}$ and $\delta_{\nu,\nu^\prime}$) for $\mu=U/2$ respectively $h=0$; in these cases the limit has to be taken carefully.

With these building blocks we can calculate all generalized susceptibilities of the HA: The $\sigma\sigma^\prime$ spin indices can be calculated directly with \cref{eq:chi_sigsig,eq:chi_sigsigp}

\begin{align}
\label{eq:chi_sigsig}
    \chi_{\sigma\sigma}^{\nu,\nu^\prime\! ,\,\omega} = G_{2,\sigma \sigma}^\text{con}\left(\nu, \nu^{\prime}, \omega\right) + \chi_{0,\sigma\sigma}^{\nu,\nu^\prime\! ,\,\omega},\\
\label{eq:chi_sigsigp}
    \chi_{\sigma\sigma^\prime}^{\nu,\nu^\prime\! ,\,\omega} = G_{2,\sigma \sigma^\prime}^\text{con}\left(\nu, \nu^{\prime}, \omega\right) \quad\text{with}\quad \sigma\neq\sigma^\prime.
\end{align}

The spin indices $\overline{\sigma\sigma^\prime}$ can be calculated with

\begin{align}
\label{eq:chi_sigbar}
    \chi_{\overline{\sigma\sigma^\prime}}^{\nu,\nu^\prime\! ,\,\omega} = - \chi_{\sigma\sigma^\prime}^{\nu,\nu+\omega ,\,\nu-\nu^\prime} + \chi_{0,\overline{\sigma\sigma^\prime}}^{\nu,\nu^\prime\! ,\,\omega},
\end{align}
where $\sigma\neq\sigma^\prime$ \cite{rohringer2013} and the bubble terms $\chi_0$ are defined in \cref{eq:chi0_ph,eq:chi0_phbar}.

Finally, the generalized susceptibilities in the particle-particle channel can be calculated by performing the frequency shift $\omega\rightarrow \omega-\nu-\nu^\prime$.

\section{Projected weight in the coupled longitudinal channel}
\label{app:weight}

To quantify how much an eigenvalue of $\chi_\mathrm{L}$ is attributed to either the charge or the spin subspace respectively, we split each eigenvector $v$ (normalized by $\norm{.}_T$ in \cref{eq:T_norm}) into the charge and spin subspaces

\begin{align}
    v = \begin{pmatrix}
        v_\text{c}\\
        v_\text{s}
    \end{pmatrix}.
\end{align}
In doing so we can assign each eigenvector a projected weight which gives the percentage contribution of the eigenvector in the charge/spin subspace $\mathcal{N}_\text{c}/\mathcal{N}_\text{s}$ with

\begin{align}
\label{eq:phys_weight}
    \mathcal{N}_\text{c/s} = \frac{\norm{v_\text{c/s}}_\dagger^2}{\norm{v}_\dagger^2},
\end{align}
where $\mathcal{N}_\text{c}+\mathcal{N}_\text{s}=1$ and  $\norm{u}_\dagger^2=u^\dagger\cdot u$\footnote{Note that for $\mathcal{N}_\text{c/s}$ other definitions would be also possible, e.g.: a) using $\norm{v}_T=1$ and defining $\mathcal{N}_\text{c/s}=\norm{v_\text{c/s}}_T$. However, since $\mathcal{N}_\text{c/s}\in\mathbb{C}$, this definition cannot be used to define a percentage. 
b)using $\norm{v}_\dagger=1$ and defining $\mathcal{N}_\text{c/s}=\norm{v_\text{c/s}}_\dagger$. This definition is in fact equivalent to the definition in \cref{eq:phys_weight}. However, vectors normalized with $\norm{.}_T$ have the convenient property of either symmetric or antisymmetric real and imaginary parts in the charge and spin subspaces (if their eigenvalue is real). Hence, it is more practical to work with these vectors. The equivalence can be shown by comparing the two different normalizations of the same vector $v$: $\tilde{v}=\frac{v^\prime}{\norm{v^\prime}_\dagger}e^{i\varphi}$, where $\norm{\tilde{v}}_\dagger=1$, $\norm{v^\prime}_T=1$, and $\varphi$ is the phase of $\norm{\tilde{v}}_T$ The additional factor $1/\norm{v}^2_\dagger$ in the definition of the projected weight assures $\mathcal{N}_\text{c}+\mathcal{N}_\text{s}= 1$.}.

To prove that pseudo-divergences with $SU(2)_P$ symmetry ($\delta\mu=0$) are always maximally ``mixed" $(\mathcal{N}_\text{c}=\mathcal{N}_\text{s}=0.5)$ we first note that the Shiba mapping connects $SU(2)_P$ symmetric case with the $SU(2)_S$ symmetric case ($h=0$), where the charge and the spin channel are decoupled centro-Hermitian matrices. 

We consider the eigenvectors $v_1$ and $v_2$ of a complex conjugate pair of eigenvalues, of the generalized susceptibility in the charge or the spin channel. Because of \cref{eq:centro_hermitian_ev} the symmetric and antisymmetric parts of $v_1$ can be written as

\begin{align}
    v_1=\frac{1}{2}(v_1+v_2^*)+\frac{1}{2}(v_1-v_2^*)\text{ with }v_\mathcal{S/A}=\frac{1}{2}(v_1\pm v_2^*).
\end{align}

As we know from \cref{eq:map_ev_even} only the symmetric part of the eigenvector gets mapped from charge/spin subspace to the spin/charge subspace.

Therefore, we must calculate the norm only for the symmetric/antisymmetric part of the eigenvector

\begin{align}
\label{eq:max_mix}
\frac{\norm{v_\mathcal{S/A}}_\dagger^2}{\norm{v_1}_\dagger^2}=\frac{1}{4\norm{v_1}_\dagger^2}(v_1^\dagger\cdot v_1\pm v_2^T\cdot v_1 \pm v_1^\dagger \cdot v_2^* + v_2^T\cdot v_2^*)=\frac{1}{2},
\end{align}
where $v_1^T\cdot v_2=0$ and $\norm{v_{2}}_\dagger=\norm{v_{1}}_\dagger$\footnote{$v_1^T\cdot v_2=0$, because the vectors are orthogonal regarding the scalar product \cref{eq:T_scalar_prod}. Further, $\norm{v_{2}}_\dagger=\norm{v_{1}}_\dagger$ holds because of the relation $v_2=Jv_1^*$}. 

Hence, we find that complex conjugate pairs, and therefore pseudo-divergences, have the maximal mixing $(\mathcal{N}_\text{c}=\mathcal{N}_\text{s}=0.5)$ between the charge and longitudinal spin subspace when $SU(2)_P$ symmetry is fulfilled. 

\section{T=0 limit of the Hubbard atom}
\label{app:T0}

The advantage of considering a \emph{dimensionless} parameter space of $\{\beta\delta\mu, \beta h, \beta U\}$ in the main text allows also for an intuitive understanding of the $T \to 0$ limit.

\begin{figure}[ht!]
\includegraphics[scale=0.2]{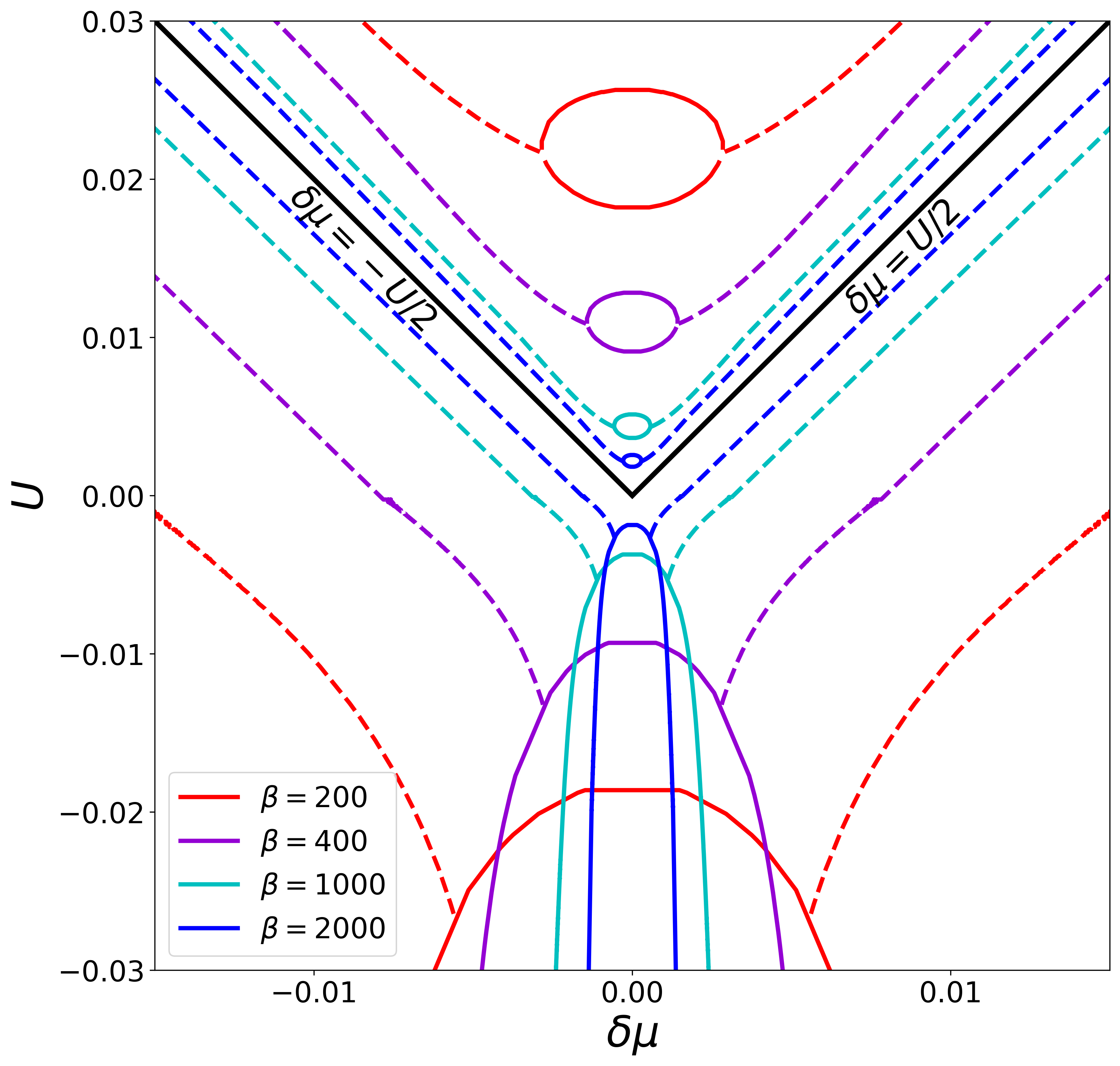}
\caption{\label{fig:T_evolv_loop} The phase space diagram of the first loop and parabola divergence line structure in the charge channel for different temperatures for the $SU(2)_S$ symmetric case ($h=0$).}
\end{figure}

In \cref{fig:T_evolv_loop} the first loop and the first parabola divergence line structure in the charge channel for the $SU(2)_S$ symmetric case are shown for different inverse temperatures $\beta=1/T$.

Here, we observe that, for $T\rightarrow 0$ ($\beta \to \infty$), the loop structure contracts to a single point and both exceptional points of the parabola approach $\delta\mu=0$.

The scaling suggested by the $T\rightarrow 0$ evolution of divergence lines shown in \cref{fig:T_evolv_loop} outlines an intuitive procedure to get a first understanding of the $T=0$ vertex divergence properties. This procedure generally works satisfactory, except for the discontinuous parts of the phase space, located at $\delta\mu=0$ for $U<0$ and at $\delta\mu=\pm U/2$ for $U>0$. In particular, since the generalized susceptibility matrix becomes Hermitian for $\delta\mu=0$ the exceptional points must transform either continuously into two degenerate or discontinuously into two separate eigenvalues of two linear-independent eigenvectors (one symmetric, one antisymmetric). 
By carefully evaluating the  analytical expressions at $\delta\mu=0$ and $T=0$ (see Refs.~\cite{thunstrom2018,essl2023}) we find a single vanishing eigenvalue with an antisymmetric eigenvector for every $U<0$ and $\delta\mu=0$ at $T=0$ where the model is discontinuous.
On the other hand, for $U>0$ where no discontinuity appears one finds two degenerate vanishing eigenvalues (one with symmetric one with antisymmetric eigenvector) for every $U>0$ and $\delta\mu=0$ at $T=0$. This behavior matches the $T\rightarrow 0$ evolution the loop structures formed by an symmetric and an antisymmetric divergence that contracts to an single point.

Further, we find that the pseudo-divergence lines (at $\delta\mu\neq0$) become parallel to $\delta\mu=\pm U/2$ in the limit of $U,\delta\mu\rightarrow \infty$ which mark the boundaries of a V-shaped region in the phase diagram. Thus, the pseudo-divergences that stem from the loop structure at $U>0$ remain inside this V-shaped region ($U>2|\delta\mu|$) and the pseudo-divergences that emerge from the parabolas at $U<0$ stay outside of it ($U<2|\delta\mu|$).

Applying the same analysis to the spin and pairing channel, one can show that the pairing divergences are confined inside the V-shaped area while the spin pseudo-divergences stay outside of it. 

From these intuitive considerations we can schematically draw \mbox{\cref{fig:divs_extreme_T0}} in the main text.

In principle, a more rigorous treatment of the $T=0$ limit, which yield however the same results~\footnote{Indeed the results obtained with the simple scaling argument of the dimensionless phase space agree well with the results from the more rigorous treatment of the $T\rightarrow0$ limit of the analytical formulas, where only special care has to be taken for the discontinuous parts of the phase space. This shows the usefulness of the dimensionless phase space.}, can be performed as explained in Ref.~\cite{essl2023}. In particular, for the analytical expressions of the generalized susceptibility reported in \cref{app:sus_HA} one has to take into account that the Matsubara frequencies become continuous and can no longer be regarded as matrix indices.

Consequently for $T=0$, a sum over Matsubara frequencies becomes an integral, and a Kronecker-Delta in Matsubara frequencies becomes a Dirac-Delta:

\begin{align}
\label{eq:T0_sum}
    \frac{1}{\beta}\sum_\nu \overset{\beta\rightarrow\infty}{\longrightarrow} \int_{-\infty}^\infty \frac{d\nu}{2\pi}, \\
\label{eq:T0_delta_nu}
    \beta \delta_{\nu\nu^\prime}\overset{\beta\rightarrow\infty}{\longrightarrow}2\pi\delta (\nu-\nu^\prime),\\
\label{eq:T0_delta_om}
    \beta \delta_{\omega 0}\overset{\beta\rightarrow\infty}{\longrightarrow}2\pi\delta (\omega).
\end{align}

For the remaining terms in the analytical formulas in \cref{app:sus_HA} we take the limit $\beta\rightarrow\infty$. 

\section{Additional information to vertex divergence structure}
\label{app:vertex}

In this section, we will give some additional information about the vertex \mbox{(pseudo-)}divergences phase diagram. 

\begin{figure}[ht!]
\includegraphics[scale=0.24]{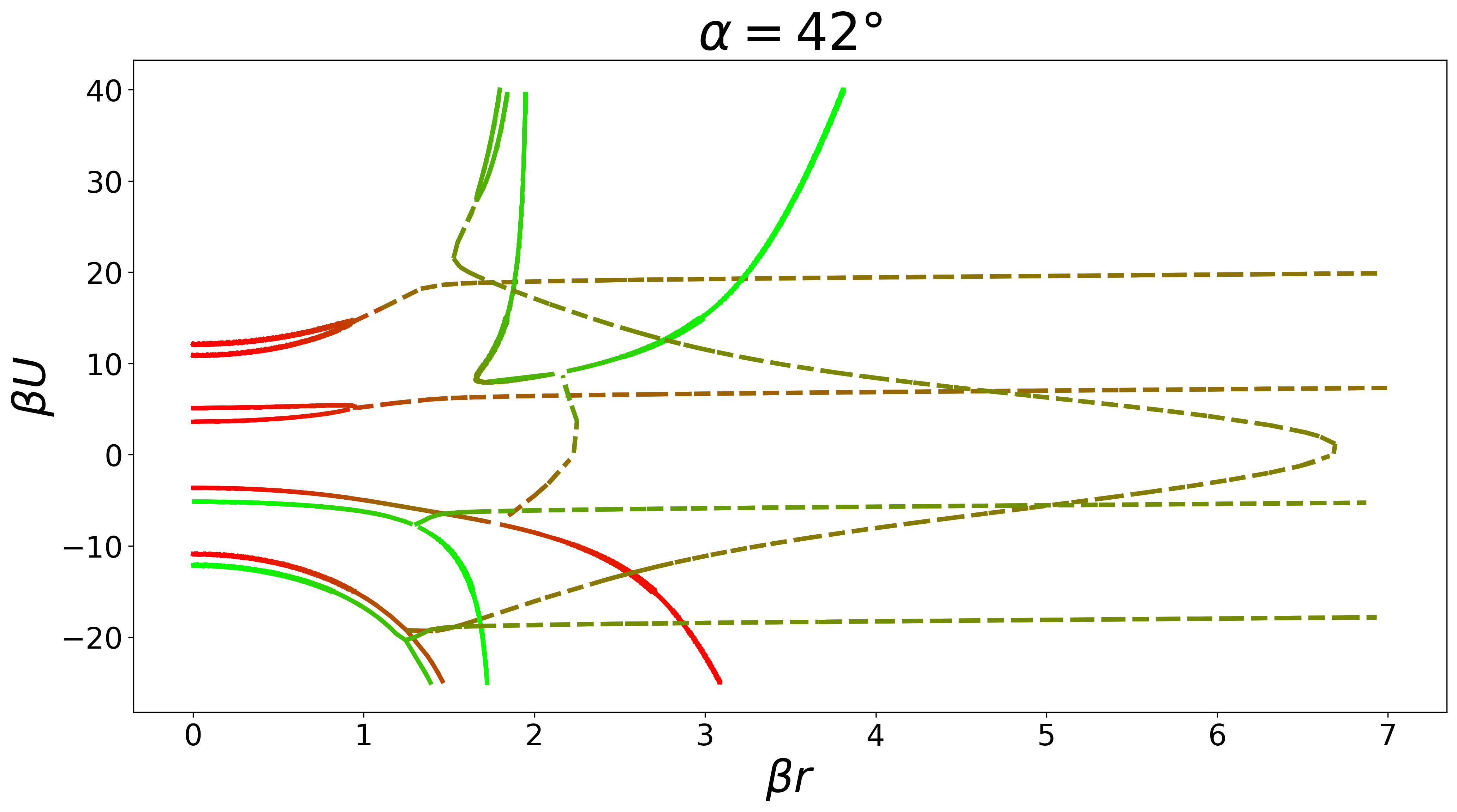}
\caption{\label{fig:more_struct}  The phase space diagrams of the vertex \mbox{(pseudo-)}divergences in the coupled longitudinal channel for the Hubbard atom are shown. Note that the first two divergence line structures are displayed. The plotting conventions are as in \cref{fig:divs_extreme}.}
\end{figure}

To get started we show the first two vertex \mbox{(pseudo-)}divergences line structures in the coupled longitudinal channel for $\alpha=42^\circ$ in \cref{fig:more_struct}. 

In addition to the first divergence line loop, we find a second loop at higher $U$ values for $U>0$.
On the other hand, for $U<0$ we find a second pair of parabolas.
Furthermore, we find a second parabola for $U>0$ at smaller $r$ values than the first parabola at $U>0$. The pseudo-divergence that is adjoined to the second parabola takes a large detour to high $r$ values and then goes back to the parabola pair at negative $U<0$.

To investigate the pseudo-divergences at $U=0$ we just have to consider the bubble term $\chi_0$ of the susceptibility.
The bubble term in the spin index space is defined as 

\begin{align}
\label{eq:chi0_ph}
    \chi_{0,\sigma\sigma^\prime}^{\nu,\nu^\prime\! ,\,\omega} = -\beta G_\sigma (\nu) G_{\sigma} (\nu+\omega) \delta_{\nu\nu^\prime} \delta_{\sigma\sigma^\prime},\\
\label{eq:chi0_phbar}
    \chi_{0,\overline{\sigma\sigma^\prime}}^{\nu,\nu^\prime\! ,\,\omega} = -\beta G_\sigma (\nu) G_{\sigma^\prime} (\nu+\omega) \delta_{\nu\nu^\prime},\\
\label{eq:chi0_pp}
    \chi_{0,\overline{\sigma\sigma^\prime},\text{pp}}^{\nu,\nu^\prime\! ,\,\omega} = -\beta G_\sigma (\nu) G_{\sigma^\prime} (\omega-\nu) \delta_{\nu\nu^\prime}
\end{align}
for the longitudinal and transversal spin channel in ph and pp frequency convention.
$\chi_0$ in the longitudinal channel reads

\begin{align}
\label{eq:chi0_l}
    \chi_{0,\text{L}} = \begin{pmatrix}
                        \chi_{0,\uparrow\uparrow}+\chi_{0,\downarrow\downarrow}& \chi_{0,\uparrow\uparrow}-\chi_{0,\downarrow\downarrow}\\
                        \chi_{0,\uparrow\uparrow}-\chi_{0,\downarrow\downarrow}& \chi_{0,\uparrow\uparrow}+\chi_{0,\downarrow\downarrow}
                        \end{pmatrix}.
\end{align}

Note that $\chi_{0,\text{L}}$ is only diagonal for $SU(2)_S$ symmetric systems, but can be diagonalized by

\begin{align}
\label{eq:chi0_sigma}
    P \chi_{0,\text{L}} P = \begin{pmatrix}
                        \chi_{0,\uparrow\uparrow}& \mathbb{0}\\
                         \mathbb{0} & \chi_{0,\downarrow\downarrow}
                        \end{pmatrix}
                        \quad\text{with}\quad P =\begin{pmatrix}
                                                                    \mathbb{1}&\mathbb{1}\\
                                                                    \mathbb{1}&-\mathbb{1}
                                                                    \end{pmatrix},
\end{align}
where $P$ transforms from the spin index basis to the longitudinal (physical) channel. Further, the bubble terms for the transversal spin and the pairing channel are defined as
\begin{align}
\label{eq:chi0_sx}
    \chi_{0,S_x} = \chi_{0,\overline{\uparrow\downarrow}} + \chi_{0,\overline{\downarrow\uparrow}},\\
\label{eq:chi0_pair}
    \chi_{0,\text{pair}} = -\chi_{0,\overline{\uparrow\downarrow},\text{pp}} -\Big(\chi_{0,\overline{\downarrow\uparrow},\text{pp}}\Big)^*.
\end{align}

By replacing $G_\sigma$ with $G_{0,\sigma}$ and by solving for Re($\chi_0$)$=0$ we find the pseudo-divergences for $U=0$ and $\omega=0$. The resulting conditions for the different channels are the following:

\begin{align}
\label{eq:pdivU0_uu}
    \chi_{0,\uparrow\uparrow}: \quad r = \frac{\nu}{\sqrt{1+ 2 \cos{\alpha}\sin{\alpha}}},
\end{align}
\begin{align}
\label{eq:pdivU0_dd}
    \chi_{0,\downarrow\downarrow}: \quad r = \frac{\nu}{\sqrt{1- 2 \cos{\alpha}\sin{\alpha}}},
\end{align}

\begin{align}
\label{eq:pdivU0_sx}
    \chi_{0,S_x}: \quad r = \frac{\nu}{\sqrt{(\cos{\alpha})^2-(\sin{\alpha})^2}},
\end{align}
\vspace{-5mm}
\begin{align}
\label{eq:pdivU0_pair}
    \chi_{0,\text{pair}}: \quad r = \frac{\nu}{\sqrt{(\sin{\alpha})^2-(\cos{\alpha})^2}}.
\end{align}

Therefore, for $\chi_{0,\uparrow\uparrow}$ and $\chi_{0,\downarrow\downarrow}$ two pseudo-divergences cross the $U=0$ axis for almost all values of $\alpha$ except for $\alpha=45^\circ$ where only the pseudo-divergences of $\chi_{0,\downarrow\downarrow}$ are present at $U=0$. 
By transforming the eigenvectors back in the space of the longitudinal channel we find that all pseudo-divergences at $U=0$ are maximally ``mixed" $(\mathcal{N}_\text{c}=\mathcal{N}_\text{s}=0.5)$.

In the transversal spin channels, pseudo-divergences cross the $U=0$ axis only for $0^\circ\leq \alpha < 45^\circ$ and for the pairing channel only for $45^\circ< \alpha \leq 90^\circ$. 

\section{Data availability}
A data set containing all numerical data and scripts for calculating and plotting these data is publicly available on the TU Wien Research Data repository \cite{dataset}.

\bibliography{refs}

\end{document}